\documentclass[10pt]{article}

\usepackage[margin=1in]{geometry}

\usepackage{graphicx}
\usepackage{hyperref}
\usepackage{booktabs}
\usepackage{tabularx}
\usepackage{soul}
\usepackage{subcaption}
\usepackage{siunitx}
\usepackage{float}
\usepackage{mathpartir}
\usepackage{amsmath}
\usepackage{amsthm}
\usepackage{algorithm}
\usepackage{algpseudocode}
\usepackage{pgf}
\usepackage[numbers]{natbib}

\newcommand{\Description}[1]{}

\theoremstyle{plain}
\newtheorem{theorem}{Theorem}[section]
\newtheorem{lemma}[theorem]{Lemma}

\theoremstyle{definition}
\newtheorem{definition}{Definition}

\theoremstyle{definition}
\newtheorem{example}{Example}

\newcommand{\subtractAndPrint}[2]{%
  \pgfmathsetmacro{\myresult}{#1 - #2}%
  \pgfmathprintnumber{\myresult}%
}

\algrenewcommand\algorithmicrequire{\textbf{Input:}}
\algrenewcommand\algorithmicensure{\textbf{Output:}}

\newcolumntype{Y}{>{\raggedleft\arraybackslash}X}

\title{Idempotent Slices with Applications to Code-Size Reduction}

\author{
Rafael Alvarenga de Azevedo\\
UFMG, Belo Horizonte, Brazil\\
\texttt{rafael.alvarenga@dcc.ufmg.br}
\and
Daniel Augusto Costa de Sa\\
UFMG, Belo Horizonte, Brazil\\
\texttt{danielaugusto191@mat-comp.grad.ufmg.br}
\and
Rodrigo Caetano Rocha\\
Huawei, Edinburgh, United Kingdom\\
\texttt{rodrigo.rocha@huawei.com}
\and
Fernando Magno Quintão Pereira\\
UFMG, Belo Horizonte, Brazil\\
\texttt{fernando@dcc.ufmg.br}
}

\begin{document}

\maketitle

\newcommand{\DaedalusInstHighCount}{62}
\newcommand{\DaedalusInstHighGeo}{0.97\%}
\newcommand{\FMInstHighCount}{59}
\newcommand{\FMInstHighGeo}{1.74\%}
\newcommand{\IROInstHighCount}{32}
\newcommand{\IROInstHighGeo}{1.54\%}
\newcommand{\DaedalusSizeHighCount}{104}
\newcommand{\DaedalusSizeHighGeo}{4.33\%}
\newcommand{\FMSizeHighCount}{156}
\newcommand{\FMSizeHighGeo}{7.51\%}
\newcommand{\IROSizeHighCount}{42}
\newcommand{\IROSizeHighGeo}{3.28\%}
\newcommand{\DaedalusExecHighCount}{93}
\newcommand{\DaedalusExecHighGeo}{11.07\%}
\newcommand{\FMExecHighCount}{101}
\newcommand{\FMExecHighGeo}{13.94\%}
\newcommand{\IROExecHighCount}{83}
\newcommand{\IROExecHighGeo}{6.47\%}
\newcommand{\DaedalusCompHighCount}{289}
\newcommand{\DaedalusCompHighGeo}{49.18\%}
\newcommand{\FMCompHighCount}{227}
\newcommand{\FMCompHighGeo}{26.85\%}
\newcommand{\IROCompHighCount}{252}
\newcommand{\IROCompHighGeo}{30.10\%}

\newcommand{\DaedalusInstLowCount}{80}
\newcommand{\DaedalusInstLowGeo}{-6.49\%}
\newcommand{\FMInstLowCount}{124}
\newcommand{\FMInstLowGeo}{-6.29\%}
\newcommand{\IROInstLowCount}{175}
\newcommand{\IROInstLowGeo}{-7.41\%}
\newcommand{\DaedalusSizeLowCount}{29}
\newcommand{\DaedalusSizeLowGeo}{-7.24\%}
\newcommand{\FMSizeLowCount}{34}
\newcommand{\FMSizeLowGeo}{-9.75\%}
\newcommand{\IROSizeLowCount}{118}
\newcommand{\IROSizeLowGeo}{-4.30\%}
\newcommand{\DaedalusExecLowCount}{68}
\newcommand{\DaedalusExecLowGeo}{-12.07\%}
\newcommand{\FMExecLowCount}{76}
\newcommand{\FMExecLowGeo}{-10.36\%}
\newcommand{\IROExecLowCount}{77}
\newcommand{\IROExecLowGeo}{-8.72\%}
\newcommand{\DaedalusCompLowCount}{14}
\newcommand{\DaedalusCompLowGeo}{-19.81\%}
\newcommand{\FMCompLowCount}{14}
\newcommand{\FMCompLowGeo}{-17.25\%}
\newcommand{\IROCompLowCount}{6}
\newcommand{\IROCompLowGeo}{-17.51\%}

\newcommand{\DaedalusInstUnchCount}{1865}
\newcommand{\DaedalusInstUnchGeo}{0.00\%}
\newcommand{\FMInstUnchCount}{1824}
\newcommand{\FMInstUnchGeo}{0.00\%}
\newcommand{\IROInstUnchCount}{1800}
\newcommand{\IROInstUnchGeo}{0.00\%}
\newcommand{\DaedalusSizeUnchCount}{1874}
\newcommand{\DaedalusSizeUnchGeo}{0.00\%}
\newcommand{\FMSizeUnchCount}{1817}
\newcommand{\FMSizeUnchGeo}{0.00\%}
\newcommand{\IROSizeUnchCount}{1847}
\newcommand{\IROSizeUnchGeo}{0.00\%}
\newcommand{\DaedalusExecUnchCount}{1846}
\newcommand{\DaedalusExecUnchGeo}{0.00\%}
\newcommand{\FMExecUnchCount}{1830}
\newcommand{\FMExecUnchGeo}{0.00\%}
\newcommand{\IROExecUnchCount}{1847}
\newcommand{\IROExecUnchGeo}{0.00\%}
\newcommand{\DaedalusCompUnchCount}{1704}
\newcommand{\DaedalusCompUnchGeo}{0.00\%}
\newcommand{\FMCompUnchCount}{1766}
\newcommand{\FMCompUnchGeo}{0.00\%}
\newcommand{\IROCompUnchCount}{1749}
\newcommand{\IROCompUnchGeo}{0.00\%}

\newcommand{\DaedalusInstTotal}{2007}
\newcommand{\DaedalusInstOverallGeo}{-0.24\%}
\newcommand{\FMInstTotal}{2007}
\newcommand{\FMInstOverallGeo}{-0.35\%}
\newcommand{\IROInstTotal}{2007}
\newcommand{\IROInstOverallGeo}{-0.65\%}

\newcommand{\DaedalusSizeTotal}{2007}
\newcommand{\DaedalusSizeOverallGeo}{0.11\%}
\newcommand{\FMSizeTotal}{2007}
\newcommand{\FMSizeOverallGeo}{0.39\%}
\newcommand{\IROSizeTotal}{2007}
\newcommand{\IROSizeOverallGeo}{-0.19\%}

\newcommand{\DaedalusExecTotal}{2007}
\newcommand{\DaedalusExecOverallGeo}{0.06\%}
\newcommand{\FMExecTotal}{2007}
\newcommand{\FMExecOverallGeo}{0.25\%}
\newcommand{\IROExecTotal}{2007}
\newcommand{\IROExecOverallGeo}{-0.09\%}

\newcommand{\DaedalusCompTotal}{2007}
\newcommand{\DaedalusCompOverallGeo}{4.22\%}
\newcommand{\FMCompTotal}{2007}
\newcommand{\FMCompOverallGeo}{2.06\%}
\newcommand{\IROCompTotal}{2007}
\newcommand{\IROCompOverallGeo}{2.48\%}

\newcommand{\DaedalusAMGmkTextSizeReduction}{-12.49\%}
\newcommand{\FMAMGmkTextSizeReduction}{-5.59\%}
\newcommand{\IROAMGmkTextSizeReduction}{-0.94\%}

\newcommand{\DaedalusSizeTextValue}{-9.32\%}
\newcommand{\FuncMergSizeTextValue}{1.07\%}
\newcommand{\IROutlinerSizeTextValue}{-0.35\%}

\newcommand{\DaedalusSizeTextValueInc}{1.97\%}
\newcommand{\FuncMergSizeTextValueInc}{-2.5\%}
\newcommand{\IROutlinerSizeTextValueInc}{-0.53\%}

\newcommand{\PosAllInstDaedalus}{14}
\newcommand{\PosAllInstFuncMerging}{26}
\newcommand{\PosAllInstIROutliner}{0}

\newcommand{\PosAllTextDaedalus}{} 
\newcommand{\PosAllTextFuncMerging}{} 
\newcommand{\PosAllTextIROutliner}{} 

\newcommand{\PosAllExecDaedalus}{} 
\newcommand{\PosAllExecFuncMerging}{} 
\newcommand{\PosAllExecIROutliner}{} 

\newcommand{\PosAllCompDaedalus}{} 
\newcommand{\PosAllCompFuncMerging}{} 
\newcommand{\PosAllCompIROutliner}{} 

\newcommand{\NegAllInstDaedalus}{0}
\newcommand{\NegAllInstFuncMerging}{0}
\newcommand{\NegAllInstIROutliner}{0}

\newcommand{\NegAllTextDaedalus}{} 
\newcommand{\NegAllTextFuncMerging}{} 
\newcommand{\NegAllTextIROutliner}{} 

\newcommand{\NegAllExecDaedalus}{}
\newcommand{\NegAllExecFuncMerging}{}
\newcommand{\NegAllExecIROutliner}{}

\newcommand{\NegAllCompDaedalus}{} 
\newcommand{\NegAllCompFuncMerging}{} 
\newcommand{\NegAllCompIROutliner}{} 

\newcommand{\NegInstTextDaedalusB}{23}
\newcommand{\NegInstTextFuncMergingB}{30}
\newcommand{\NegInstTextIROutlinerB}{105}

\newcommand{\NegTextDaedalusB}{} 
\newcommand{\NegTextFuncMergingB}{} 
\newcommand{\NegTextIROutlinerB}{} 

\newcommand{\DiffGeoNegInstDaedalusB}{-8.39\%}
\newcommand{\DiffGeoNegInstcountDaedalusB}{-9.96\%}
\newcommand{\DiffGeoNegInstFuncMergingB}{-10.72\%}
\newcommand{\DiffGeoNegInstcountFuncMergingB}{-8.09\%}
\newcommand{\DiffGeoNegInstIROutlinerB}{-4.65\%}
\newcommand{\DiffGeoNegInstcountIROutlinerB}{-7.58\%}

\newcommand{\PosInstTextDaedalus}{57}
\newcommand{\PosInstTextFuncMerging}{56}
\newcommand{\PosInstTextIROutliner}{10}

\newcommand{\PosTextDaedalus}{} 
\newcommand{\PosTextFuncMerging}{} 
\newcommand{\PosTextIROutliner}{} 

\newcommand{\DiffGeoPosInstDaedalus}{2.09\%}
\newcommand{\DiffGeoPosInstFuncMerging}{6.02\%}
\newcommand{\DiffGeoPosInstIROutliner}{3.37\%}

\newcommand{\OriginalLdecod}{3007}
\newcommand{\DaedalusLdecod}{3052}
\newcommand{\CreatedFuncADaedLdecod}{9}
\newcommand{\CreatedFuncBDaedLdecod}{9}
\newcommand{\IROutlinerLdecod}{3200}
\newcommand{\CreatedFuncAIROLdecod}{174}
\newcommand{\CreatedFuncBIROLdecod}{180}
\newcommand{\FMLdecod}{2963}
\newcommand{\CreatedFuncAFMLdecod}{211}
\newcommand{\CreatedFuncBFMLdecod}{514}
\newcommand{\CreatedFuncCFMLdecod}{303}
\newcommand{\CreatedFuncDFMLdecod}{314}
\newcommand{\CreatedFuncEFMLdecod}{515}
\newcommand{\CreatedFuncFFMLdecod}{1145}
\newcommand{\FMNewFuncsNumCalls}{24}
\newcommand{\IRONewFuncsNumCalls}{9}
\newcommand{\DaedNewFuncsNumCalls}{2}

\newcommand{\NegInstPosExecDaedalus}{14}
\newcommand{\NegInstPosExecFuncMerging}{6}
\newcommand{\NegInstPosExecIROutliner}{29}

\newcommand{\NegTextPosExecDaedalus}{} 
\newcommand{\NegTextPosExecFuncMerging}{} 
\newcommand{\NegTextPosExecIROutliner}{} 

\newcommand{\PosExecTimeDaedalus}{} 
\newcommand{\PosExecTimeFuncMerging}{} 
\newcommand{\PosExecTimeIROutliner}{} 

\newcommand{\DiffGeoPosExecDaedalus}{4.48\%}
\newcommand{\DiffGeoPosExecFuncMerging}{19.96\%}
\newcommand{\DiffGeoPosExecIROutliner}{6.67\%}

\newcommand{\NegInstExecDaedalus}{7}
\newcommand{\NegInstExecFuncMerging}{6}
\newcommand{\NegInstExecIROutliner}{27}

\newcommand{\NegTextExecDaedalus}{} 
\newcommand{\NegTextExecFuncMerging}{} 
\newcommand{\NegTextExecIROutliner}{} 

\newcommand{\NegExecTimeDaedalus}{} 
\newcommand{\NegExecTimeFuncMerging}{} 
\newcommand{\NegExecTimeIROutliner}{} 

\newcommand{\DiffGeoExecDaedalus}{-3.39\%}
\newcommand{\DiffGeoExecFuncMerging}{-15.16\%}
\newcommand{\DiffGeoExecIROutliner}{-8.34\%}

\newcommand{\NegCompTimeDaedalus}{9}
\newcommand{\NegCompTimeFuncMerging}{11}
\newcommand{\NegCompTimeIROutliner}{5}

\newcommand{\DiffGeomCompDaedalus}{-19.81\%}
\newcommand{\DiffGeomCompFuncMerging}{-17.25\%}
\newcommand{\DiffGeomCompIROutliner}{-17.51\%}

\newcommand{\NegInstDaedalus}{0}
\newcommand{\NegInstFuncMerging}{1}
\newcommand{\NegInstIROutliner}{1}

\newcommand{\NegTextDaedalus}{} 
\newcommand{\NegTextFuncMerging}{} 
\newcommand{\NegTextIROutliner}{} 

\newcommand{\NegCompDaedalus}{} 
\newcommand{\NegCompFuncMerging}{} 
\newcommand{\NegCompIROutliner}{} 

\newcommand{\DiffGeoDaedalus}{-}
\newcommand{\DiffGeoFuncMerging}{-8.57\%}
\newcommand{\DiffGeoIROutliner}{-1.61\%}

\newcommand{\DaedalusInstcount}{23}
\newcommand{\FuncMergingInstcount}{19}
\newcommand{\IROutlinerInstcount}{82}

\newcommand{\DaedalusTextsize}{} 
\newcommand{\FuncMergingTextsize}{} 
\newcommand{\IROutlinerTextsize}{} 

\newcommand{\DaedalusComptime}{} 
\newcommand{\FuncMergingComptime}{} 
\newcommand{\IROutlinerComptime}{} 

\newcommand{\DaedalusDiffGeomean}{80.65\%}
\newcommand{\FuncMergingDiffGeomean}{24.31\%}
\newcommand{\IROutlinerDiffGeomean}{22.35\%}

\newcommand{\DaedalusCorr}{0.837392779}
\newcommand{\FuncMergingCorr}{0.940710506}
\newcommand{\IROutlinerCorr}{0.918238062}

\newcommand{\dedirofumInstcount}{-1.22\%}
\newcommand{\fumdediroInstcount}{-1.01\%}
\newcommand{\dedfumiroInstcount}{-1.07\%}
\newcommand{\fumirodedInstcount}{-0.98\%}
\newcommand{\irodedfumInstcount}{-1.19\%}
\newcommand{\irofumdedInstcount}{-1.16\%}

\newcommand{\dedirofumText}{0.43\%}
\newcommand{\fumdediroText}{0.33\%}
\newcommand{\dedfumiroText}{0.38\%}
\newcommand{\fumirodedText}{0.33\%}
\newcommand{\irodedfumText}{0.41\%}
\newcommand{\irofumdedText}{0.39\%}

\newcommand{\dedirofumExec}{0.28\%}
\newcommand{\fumdediroExec}{0.28\%}
\newcommand{\dedfumiroExec}{0.37\%}
\newcommand{\fumirodedExec}{0.39\%}
\newcommand{\irodedfumExec}{0.16\%}
\newcommand{\irofumdedExec}{0.29\%}

\newcommand{\dedirofumComp}{5.57\%}
\newcommand{\fumdediroComp}{5.31\%}
\newcommand{\dedfumiroComp}{5.49\%}
\newcommand{\fumirodedComp}{5.20\%}
\newcommand{\irodedfumComp}{5.50\%}
\newcommand{\irofumdedComp}{5.09\%}

\begin{abstract}
Given a value computed within a program, an idempotent backward slice with respect to this value is a maximal subprogram that computes it.
An informal notion of an idempotent slice has previously been used by Guimar\~{a}es et al. to transform eager into strict evaluation in the LLVM intermediate representation.
However, that algorithm is insufficient to be correctly applied to general control-flow graphs.
This paper addresses these omissions by formalizing the notion of idempotent backward slices and presenting a sound and efficient algorithm for extracting them from programs in Gated Static Single Assignment (GSA) form.
As an example of their practical use, the paper describes how identifying and extracting idempotent backward slices enables a sparse code-size reduction optimization;
that is, one capable of merging non-contiguous sequences of instructions within the control-flow graph of a single function or across functions.
Experiments with the LLVM test suite show that, in specific benchmarks, this new algorithm achieves code-size reductions up to -7.24\% on programs highly optimized by the -Os sequence of passes from clang 17.
\end{abstract}


\section{Introduction}
\label{sec_intro}

Recently, \citet{campos2023} introduced a technique to convert call-by-value evaluation into call-by-need.
Their approach computes, in a backward manner, the dependencies of a function parameter $v$, outlines the corresponding subprogram into a separate function, and invokes this new function within the callee whenever $v$ is used.
In this paper, we call such a subprogram an {\it idempotent backward slice} with criterion $v$.
Idempotent backward slices have applications beyond ``lazyfication.''
For instance, outlining such slices enables hot-cold code splitting, reveals task-level parallelism, serves as a mechanism for code obfuscation, and supports code-size reduction, as recurring slices can be consolidated into a single function.

We have observed that \citeauthor{campos2023}'s algorithm fails to identify slices in at least two scenarios.
First, as already indicated by the authors themselves, their approach may fail to extract slices from programs that do not satisfy the ``conventional'' Static Single Assignment (CSSA) property~\cite{Sreedhar99}, such as those containing variables related by $\phi$-functions whose live ranges overlap.
Second, as shown in Section~\ref{sub_problemWyvern}, the algorithm may fail to extract slices from specific control-flow graphs that do not exhibit a ``hammock'' structure~\cite{Ferrante87}; that is, graphs that cannot be decomposed into a tree of single-entry-single-exit regions.
This paper provides an algorithm that identifies and extracts idempotent backward slices in both situations.

\paragraph{Sound Idempotent Backward Slices.}
In Section~\ref{sec_slices}, we formalize the notion of idempotent backward slices and introduce a sound algorithm to extract them as standalone functions.
The key to soundness is the observation that correct idempotent backward slices can be extracted from programs in the {\it Gated Static Single Assignment} (GSA) form, provided that GSA is built using the path-numbering algorithm proposed by \citet{tu1995}.
The identification of a single slice runs in linear time with respect to the number of edges in the program's control-flow graph and requires neither the hammock structure nor the CSSA property.

As an application of idempotent backward slices, Section~\ref{sec_size} demonstrates how they can be used to enable code-size reduction.
The idea of this optimization is to identify as many slices as possible, outline those that are isomorphic, and merge them into a single function, reducing code replication.
In contrast to previous work, such as \citet{rocha2020}'s function merging by sequence alignment or the LLVM \texttt{IROutliner} pass~\cite{Riddle2017Interprocedural}, the proposed optimization can merge non-contiguous or non-ordered instructions, or groups of instructions within the same function.

\paragraph{Summary of Contributions. }
We implemented the slicing algorithm described in Section~\ref{sec_slices} within the LLVM compiler, version 17.0.6.
This implementation is sufficiently robust to handle the entire LLVM test suite, which comprises \text{\DaedalusSizeTotal~programs}.
The experiments in Section~\ref{sec_eval} show that the code-size reduction algorithm derived from this implementation can reduce the \texttt{.text} section size from \text{\DaedalusSizeLowCount~programs}, with a geometric mean of \DaedalusSizeLowGeo.
These improvements are not subsumed by any publicly available code-size reduction techniques, such as \citet{rocha2020}'s function merging or the LLVM \texttt{IROutliner}.
Conversely, our implementation does not subsume these techniques either; they are complementary, each achieving unique code-size reductions.
For instance, our implementation achieved its greatest benefit on the \texttt{AMGmk} benchmark, reducing the \texttt{.text} section size by \text{\DaedalusAMGmkTextSizeReduction~on} top of \texttt{clang -Os}.
By comparison, \citeauthor{rocha2020}'s approach achieved a code size difference of \FMAMGmkTextSizeReduction, while the LLVM outliner achieved only \IROAMGmkTextSizeReduction.

\section{On the Challenge of Computing Sparse Idempotent Slices}
\label{sec_ovf}

The goal of this section is twofold: first, to introduce the core definitions used throughout this paper; and second, to discuss why previous work can fail to correctly identify idempotent backward slices in some cases.

\subsection{Idempotent Executions}
\label{sub_idempotent}

This paper is concerned about sequences of instructions that implement
``{\it Idempotent Executions}'', a notion that Definition~\ref{def_idempotent}
formalizes, and Example~\ref{ex_idempotent} illustrates in the context of the
LLVM compilation infrastructure:

\begin{definition}[Idempotent Execution]
\label{def_idempotent}
Let $P$ be a program in Static Single Assignment (SSA) form, and let $S$ be a
set of instructions of $P$. Let $I$ be the set of variables that are used in
$S$ but defined outside $S$ (the \emph{inputs} of $S$). The execution of $S$ is
\emph{idempotent} if, for any binding of values to the variables in $I$,
executing the instructions of $S$ multiple times produces the same values for
all variables defined in $S$ and does not change the observable state of the
program.
\end{definition}

\begin{figure}[ht]
  \centering
  \includegraphics[width=1.0\textwidth]{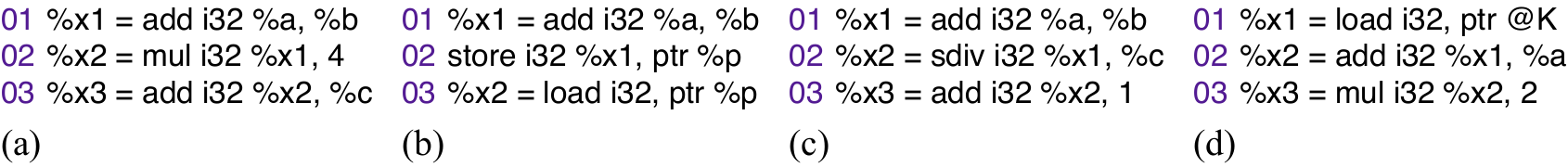}
  \caption{Examples of idempotent and non-idempotent instruction sequences in LLVM IR.}
  \Description{Examples of idempotent and non-idempotent instruction sequences in LLVM IR.}
  \label{fig_idempotent_sequences}
\end{figure}

\begin{example}
\label{ex_idempotent}
Figure~\ref{fig_idempotent_sequences} illustrates sequences of LLVM instructions that
either satisfy or violate the notion of idempotent execution. The sequence in
Fig.~\ref{fig_idempotent_sequences}(a) is idempotent because it contains only pure
arithmetic operations: executing these instructions multiple times with the
same bindings for the free variables always produces the same values and does
not modify the program state.
The sequence in Fig.~\ref{fig_idempotent_sequences}(b) is not idempotent because it
contains a \texttt{store} instruction, which modifies memory. Re-executing this
sequence may therefore change the observable state of the program.
The fragment in Fig.~\ref{fig_idempotent_sequences}(c) is also not idempotent because the
division instruction may raise an exception (for instance, if \texttt{\%c} is
zero). Instructions that may raise exceptions cannot be safely re-executed
without potentially altering the control flow of the program.
Finally, the sequence in Fig.~\ref{fig_idempotent_sequences}(d) contains a \texttt{load}
instruction but can still be considered idempotent because the value is loaded
from a constant global variable. Since the memory location \texttt{@K} is
immutable, repeated executions of the sequence always read the same value and
do not change the program state.
\end{example}

Section~\ref{sec_eval} evaluates an implementation of a compiler pass that
identifies idempotent sequences of instructions that form program slices, a
notion that we formalize in Section~\ref{sub_backward_slice}. To identify such
sequences, our implementation adopts a best-effort approach. In particular, we
exclude from an idempotent set any instruction that may raise exceptions, call
functions, or write to memory. We also exclude load instructions that may read
from mutable memory locations. However, we allow load instructions that read
from immutable memory, such as constant global variables, as seen in Figure~\ref{fig_idempotent_sequences}(d).

\subsection{Idempotent Backward Slices}
\label{sub_backward_slice}

A \emph{program slice} is a subset of a program's statements that may affect a specific point in the program, called the \emph{slice criterion}.
Early applications of program slicing, as proposed by \citet{Weiser81}, aimed to help developers focus on relevant code during tasks such as debugging and testing.
Later, the concept of slicing proved useful in compiler optimizations such as auto-parallelization~\cite{Muller01}, redundancy elimination~\cite{Tip94}, and the removal of cold code from critical paths~\cite{Wagner11}.

The standard formalization of program slices, as presented by \citet{Blazy15}, refers to \emph{dense slices}.
Given a program $P$, a dense slice is a subprogram $S \subseteq P$ that determines the computation of an instruction $\iota \in P$.
In contrast, a \emph{sparse slice}, as proposed by \citet{Rodrigues16}, captures the computation of a \emph{value} in a program represented in \emph{Static Single Assignment (SSA)} form~\cite{cytron1991}.
This paper focuses on a variation of sparse slices that we call \emph{idempotent slices}, defined as follows:

\begin{definition}[Idempotent Backward Slice]
\label{def_slice}
Let $v$ be a variable defined in an SSA-form program $P$.
The idempotent backward slice $S \subseteq P$ with respect to $v$, the slice criterion, is a subset of instructions in $P$ satisfying the following properties:
\begin{description}
\item[Single-Entry:] The set of basic blocks $B_S \subseteq P$ that contain instructions in $S$
forms a single-entry region in $P$; that is, there exists a unique basic block
$b \in B_S$ such that every path from the entry of $P$ to any block in $B_S$
must first pass through $b$.
\item[Idempotent:] Let $I$ be the set of free variables in $S$ (its inputs). Multiple executions of $S$ with the same binding of values to variables of $I$ always yield the same result (see Definition~\ref{def_idempotent}).
\end{description}
\end{definition}

Example~\ref{ex_exampleSlices} illustrates some of the concepts introduced in Definition~\ref{def_slice}.
The example mentions the Gated Static Single-Assignment form, which Section~\ref{sub_gsa} will revist in more detail.

\begin{figure}[ht]
  \centering
  \includegraphics[width=1.0\textwidth]{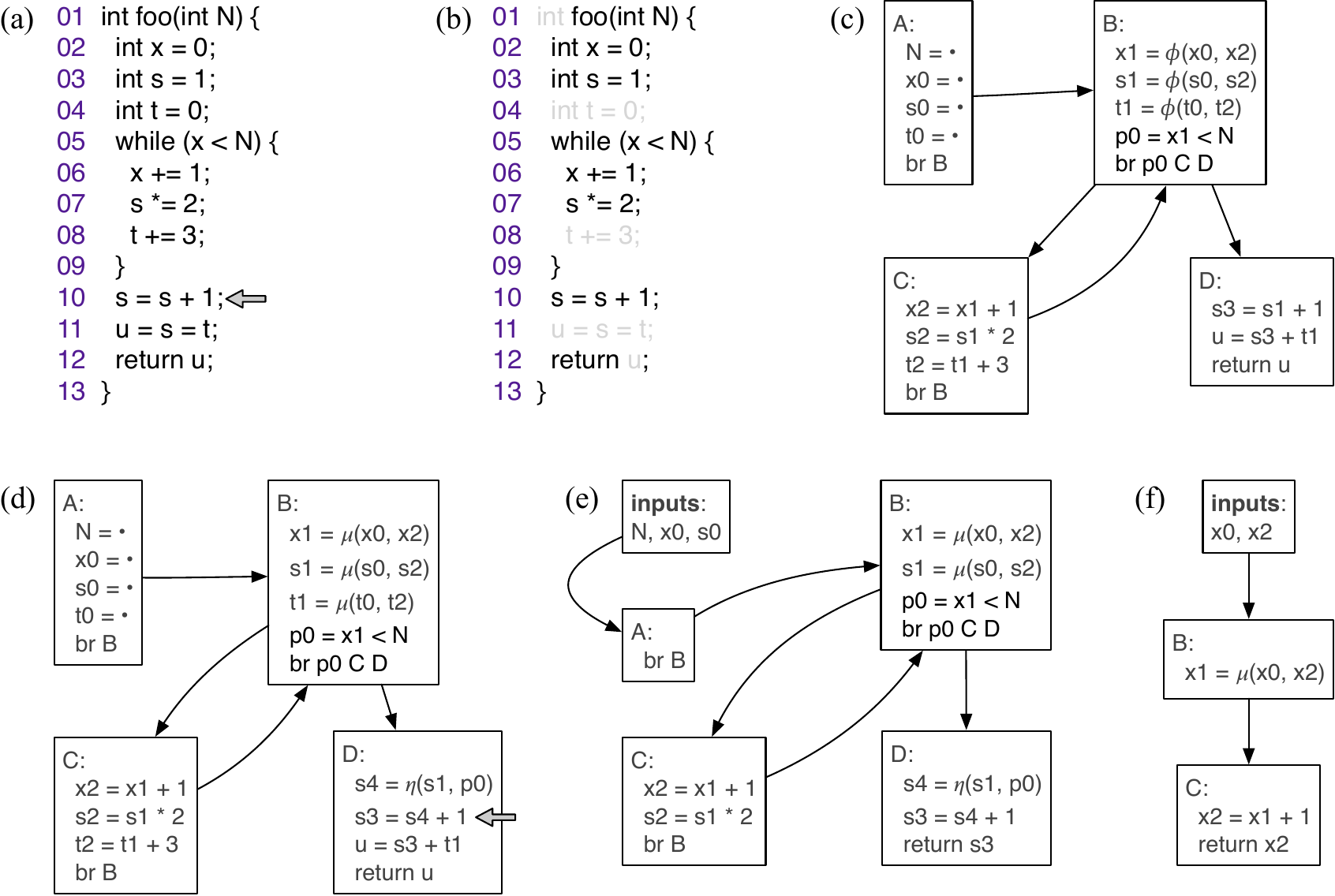}
  \caption{(a) Program with a slice criterion: the instruction $s = s + 1$.
  (b) Dense slice, adapted from \citet[Fig.~1]{Blazy15}.
  (c) Program in Static Single-Assignment form.
  (d) Program in Gated Static Single-Assignment form.
  (e) Idempotent backward slice with respect to SSA variable $s_3$.
  (f) Idempotent backward slice with respect to SSA variable $x_2$.}
  \Description{Examples of different program slices.}
  \label{fig_exampleSlices}
\end{figure}

\begin{example}
\label{ex_exampleSlices}
We use the program in Figure~\ref{fig_exampleSlices} (a) to illustrate the notion of idempotent slices.
The semantics of this program is immaterial for the discussion that follows.
Figure~\ref{fig_exampleSlices} (b), adapted from Figure~\ref{fig_exampleSlices} (e) in the work of \citet{Blazy15}, shows a dense slice computed for the instruction $s = s + 1$ in line~10.
Figure~\ref{fig_exampleSlices} (c) shows the control-flow graph of the same program converted to SSA form.
Part (d) of the figure depicts the corresponding Gated Static Single-Assignment representation, in which $\phi$-functions at loop headers are replaced with $\mu$-functions.
Additionally, a $\eta$-function renames variable $s_1$ into $s_4$ as it exits the loop.
Note that the $\eta$-function is guarded by predicate $p_0$, which controls the loop.
Figure~\ref{fig_exampleSlices} (e) shows the backward idempotent slice with respect to variable $s_3$, and Figure~\ref{fig_exampleSlices} (f) shows the slice with respect to the criterion $x_2$.
The value of $s_3$ in Figure~\ref{fig_exampleSlices} (e) depends exclusively of the set of inputs $\{N, x_0, s_0 \}$.
Notice that $s_3$ is defined outside the loop; hence, regardless of this loop's trip count, $s_3$ is assigned only once.
Similarly, the value of $x_2$ in Figure~\ref{fig_exampleSlices} (f) depends only from the initial bindings of the inputs $x_0$ and $x_2$.
Notice that, in this case, the backward slice does not escape the loop that encloses the slice criterion, otherwise $x_2$ would be affected multiple times.
\end{example}

Example~\ref{ex_exampleSlices} highlights an important distinction between the classic, i.e., dense, notion of a backward slice and the \emph{idempotent backward} slice introduced in this paper.
A dense backward slice, as defined in the original work of \citet{Weiser81} and formalized by \citet{Blazy15}, is a subset of \emph{labels} in a program that affect the slice criterion in the same way as the original program.
An idempotent slice, in contrast, is a referentially transparent function that defines a single \emph{value}.
As a result, a maximal backward slice must be restricted to the loop in which its slice criterion is defined; otherwise, it would compute multiple values for that criterion: one per loop iteration.
For this reason, the slice of $x_2$ in Figure~\ref{fig_exampleSlices} (f) contains only two basic blocks and never escapes the loop where $x_2$ is defined.

\subsection{Gated Static Single-Assignment Form}
\label{sub_gsa}

The \emph{Gated Static Single Assignment} form~\cite{tu1995,ottenstein1990} mentioned in Example~\ref{ex_exampleSlices} extends the conventional Static Single Assignment representation~\cite{cytron1991} by making explicit the control conditions that govern the flow of values through a program.
In SSA, $\phi$-functions merge values from multiple predecessors but omit the control predicates that determine which value is chosen at runtime.
GSA replaces these $\phi$-functions with three kinds of \emph{gate instructions}—$\gamma$, $\mu$, and $\eta$—which jointly encode data and control dependencies:
\begin{itemize}
\item \textbf{$\mu$-instructions:} represent \emph{loop headers} in GSA form.  
They replace $\phi$-functions when the join point corresponds to a loop entry.
A $\mu$-instruction expresses the recurrence of values across iterations by associating each incoming value with the control predicate that enables the corresponding loop back edge.
Hence, $\mu$-instructions make explicit the control dependencies governing value flow in iterative constructs.

\item \textbf{$\gamma$-instructions:} represent \emph{simple junction points}.
They generalize $\phi$-functions by pairing each incoming value with a \emph{gate}—a predicate characterizing the control-flow path from the reaching definition to the merge point.
At runtime, the argument whose predicate evaluates to true is selected.
Thus, $\gamma$-instructions encode more information than $\phi$-functions, as they explicitly represent both control and data dependencies.

\item \textbf{$\eta$-instructions:} represent \emph{value gating}.
They propagate values only under certain control conditions, linking a value's liveness to a control predicate.
In Figure~\ref{fig_exampleSlices} (d), $\eta$-instructions ensure that values produced within loops or conditional regions are properly guarded before being used or returned outside those regions.
\end{itemize}

The GSA representation enables analyses and transformations that require joint reasoning about data and control flow.
As Section~\ref{sec_slices} will discuss, one such application is the computation of idempotent backward slices.

\subsection{Limitations of Previous Work to Compute Idempotent Slices}
\label{sub_problemWyvern}

In 2023, \citet{campos2023} proposed a fast algorithm to identify backward slices.
Their goal was to transform function arguments into \emph{thunks}; that is, lambda expressions that are evaluated lazily, only when needed by the callee.
To produce a thunk encoding the computation of an actual parameter $p$, \citeauthor{campos2023} outline the backward slice that calculates the value of $p$.
In order to identify these slices efficiently, they traverse the {\it sparse} dependence graph that produces the value $p$.
For completeness, Definition~\ref{def_dep_graph} formalizes the notion of a sparse backward dependence graph.

\begin{definition}[Sparse Backward Dependence Graph]
\label{def_dep_graph}
The dependence graph $G = (V, E)$ relative to the SSA variable $v$ with program $P$ is defined as follows:
\begin{description}
\item [Base:] Variable $v \in V$;
\item [Data:] If $B: v_0 = v_1 \oplus v_2$ is an instruction in basic block $B \in P$, and $v_0 \in V$, then $\{v_1, v_2\} \subset V$, and $\{v_1 \rightarrow v_0, v_2 \rightarrow v_0\} \subseteq E$;
\item [Control:] If $B: \mathit{br} \ p \ B_0 \ B_1$ is a branch that terminates basic block $B \in P$, then for every instruction $B': v_0 = \ldots$ in basic block $B' \in P$, such that: (i) $B$ dominates $B'$; (ii) $B'$ does not dominate $B$; and (iii) $v_0 \in V$, we have that $p \in V$ and $p \rightarrow v_0 \in E$.
\end{description}
\end{definition}

\begin{figure}[ht]
  \centering
  \includegraphics[width=1.0\textwidth]{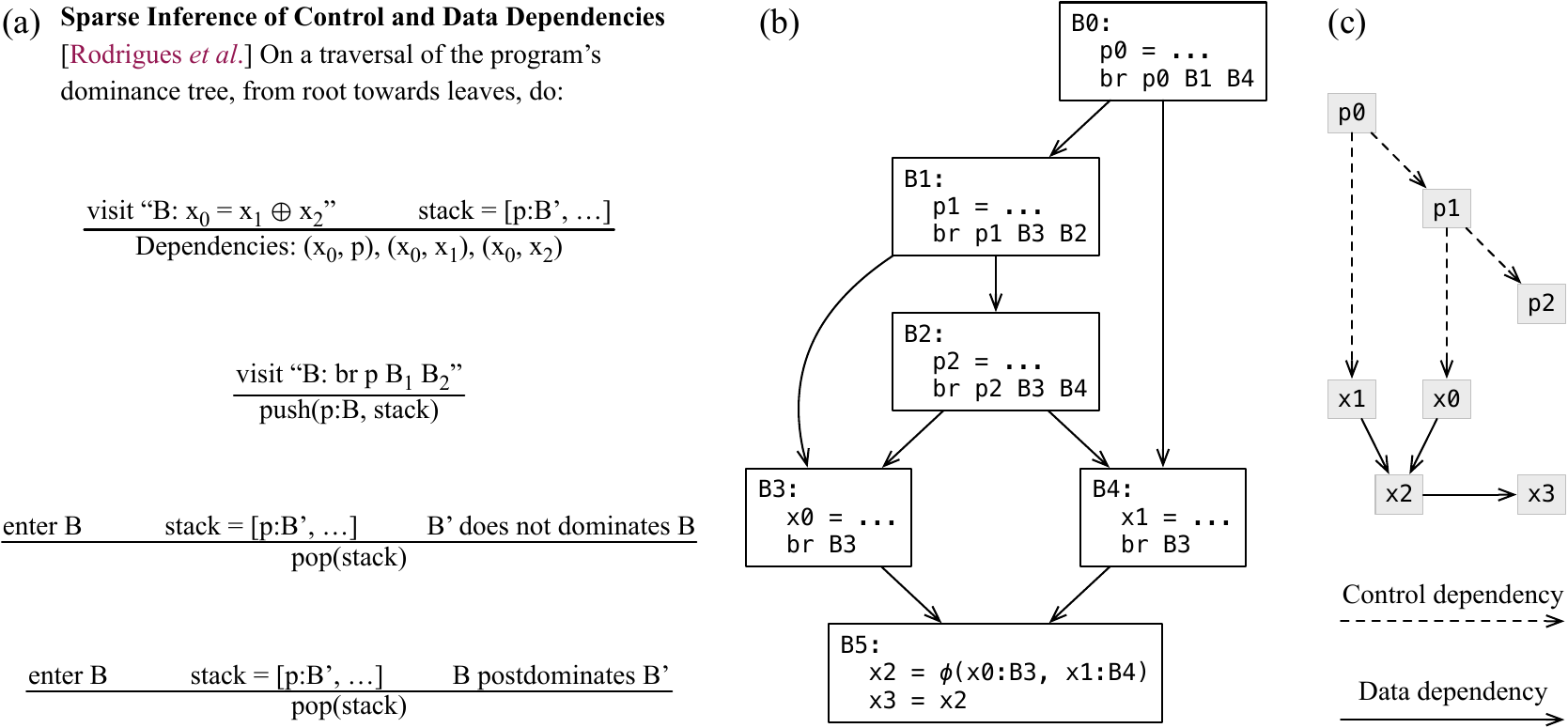}
  \caption{(a) The four rules used by \citet{Rodrigues16} to infer data and control dependencies.
  (b) A control-flow graph showing a sub-structure that causes \citet{campos2023}'s algorithm to miss block $B_2$ in the backward slice with regards to criterion $x_3$. (c) The backward dependency graph built for the slice criterion $x_3$.}
  \Description{Problem with the Wyvern approach.}
  \label{fig_thePhrygianProblem}
\end{figure}

\citeauthor{campos2023} uses the sparse method introduced by \citet{Rodrigues16} to compute the backward dependence graph of Definition~\ref{def_dep_graph}.
This algorithm, in turn, is based on the definition of control dependence by \citet{Ferrante87}:  
a basic block $B_s$ \emph{controls} a basic block $B_u$ if $B_s$ dominates $B_u$, but $B_u$ does not post-dominate $B_s$.
From this observation, \citeauthor{Rodrigues16} uses the four rules seen in 
Figure~\ref{fig_thePhrygianProblem} (a) to infer data and control dependencies.
The algorithm of \citeauthor{Rodrigues16} traverses the program's dominator tree, stacking predicates that control branches.
Upon visiting instructions that define new values, it creates control dependencies between the predicate on the top of the stack and such values.
However, when this notion, originally defined for program labels, is transposed to SSA-form variables, \citeauthor{Rodrigues16}'s algorithm misses certain control-dependency edges.
In our reimplementation of \citeauthor{campos2023}'s approach, these missing edges proved necessary to correctly capture all branches in the extracted slice, as illustrated in Example~\ref{ex_problemWyvern}.

\begin{example}
\label{ex_problemWyvern}
Figure~\ref{fig_thePhrygianProblem} (b) shows a simplified control-flow pattern that emerged from the compilation of function \texttt{reduceDB}, from the \texttt{minisat} benchmark from the LLVM test suite\footnote{Available at \url{https://github.com/llvm/llvm-test-suite/blob/main/MultiSource/Applications/minisat/Solver.cpp}} once it is optimized with the \texttt{simplifycfg} pass from the \texttt{-Os} flag.
Figure~\ref{fig_thePhrygianProblem} (c) shows the backward dependency graph with regards to criterion \texttt{x3} that the algorithm proposed by \citet{Rodrigues16} builds.
When producing the slice for \texttt{x3}, \citeauthor{campos2023}'s algorithm would miss the basic block \texttt{B2}.
This happens because, according to \citet{Ferrante87}'s definition of control dependencies, \texttt{B2} does not control neither \texttt{B3} or \texttt{B4}, nor \texttt{B5}, as it does not dominate any of these nodes.
Thus, when following this definition of control dependency, \citeauthor{Rodrigues16}'s algorithm misses the fact that predicate \texttt{p2} also controls the assignments into \texttt{x0} and \texttt{x1}.
\end{example}

The pattern that Figure~\ref{fig_thePhrygianProblem} (b) illustrates is very specific: we could identify it only once among \text{\DaedalusSizeTotal~programs} in the LLVM test suite.
Nevertheless, its occurrence leads to an incorrect computation of backward dependencies, following \citeauthor{Rodrigues16}'s algorithm.
Section~\ref{sec_slices} shall present a new algorithm to compute sparse backward slices that is correct under this scenario.

\section{Idempotent Backward Slices}
\label{sec_slices}

This section presents an algorithm to identify idempotent backward slices that meet the properties listed in Definition~\ref{def_slice}, and that overcome the limitations discussed in Section~\ref{sub_problemWyvern}.
We start explaining, in Section~\ref{sub_identification}, how such slices can be correctly identified.
Then, in Section~\ref{sub_outlining}, we show how the slice can be outlined as a stand-alone function.

\subsection{From SSA to GSA}
\label{sub_ssa_to_gsa}

To build the GSA representation, we follow the algorithm of \citet{tu1995}, which adds gates (predicates) to SSA-form programs.
This conversion unifies the placement of $\phi$-functions and the derivation of the control predicates that guard their incoming values into a single, almost-linear-time process based on the concept of a \emph{gating path}.
Given a $\phi$-function located at a merge block $B$, a gating path with respect to this $\phi$-function satisfies the following informal properties:
\begin{enumerate}
\item It starts at the immediate dominator $\text{idom}(B)$.
The immediate dominator\footnote{
A basic block $d$ \emph{dominates} a block $n$ if every path from the program entry to $n$ goes through $d$.
Dually, $d$ \emph{post-dominates} $n$ if every path from $n$ to the program exit goes through $d$.
Dominance and post-dominance form trees: the \emph{immediate dominator} (resp.\ \emph{immediate post-dominator}) of a block is its unique strict dominator (resp.\ strict post-dominator) that is closest to it in the control-flow graph.
} of $B$ is the unique point where control flow last converges before diverging toward $B$.
\item It corresponds to a unique sequence of branch predicates that determine which reaching definition flows into $B$.
\item It ends at $B$, the block containing the $\phi$-function.
\end{enumerate}

At merge points, $\phi$-functions are thus replaced by \emph{$\gamma$-functions}, which explicitly encode the Boolean predicates characterizing their gating paths.
Example~\ref{ex_gating_path} illustrates this notion.

\begin{figure}[htb]
  \centering
  \includegraphics[width=1.0\textwidth]{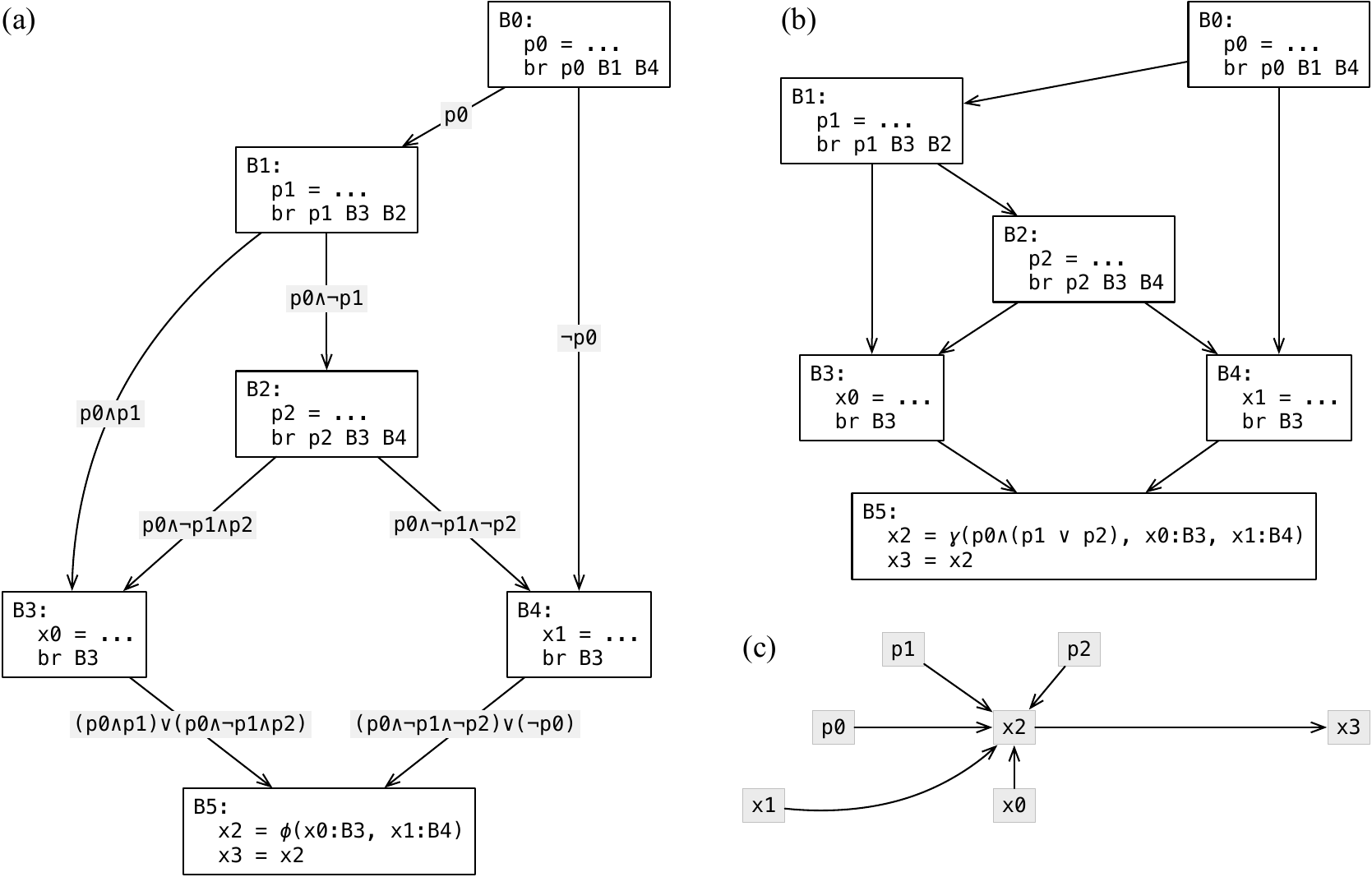}
  \caption{(a) SSA CFG annotated with gating-path predicates.
  (b) GSA form after $\phi$-elimination.
  (c) Backward dependence graph of variable \texttt{x3}.}
  \Description{Example of GSA conversion with explicit gating-path conditions.}
  \label{fig_ssa_to_gsa}
\end{figure}

\begin{example}
\label{ex_gating_path}
Consider Figure~\ref{fig_ssa_to_gsa}(a).
The merge block \texttt{B5} contains a $\phi$-function for $\mathtt{x2}$.
Its immediate dominator is \texttt{B0}, so every gating path must originate at \texttt{B0} and terminate at \texttt{B5}.
From the control-flow structure, we enumerate four distinct gating paths:

\begin{itemize}
  \item Path 1 ($\mathtt{x0}$):
  $\texttt{B0} \to \texttt{B1} \to \texttt{B3} \to \texttt{B5}$  
  Condition: $\mathtt{p0} \land \mathtt{p1}$
  \item Path 2 ($\mathtt{x0}$):
  $\texttt{B0} \to \texttt{B1} \to \texttt{B2} \to \texttt{B3} \to \texttt{B5}$  
  Condition: $\mathtt{p0} \land \neg \mathtt{p1} \land \mathtt{p2}$
  \item Path 3 ($\mathtt{x1}$):
  $\texttt{B0} \to \texttt{B1} \to \texttt{B2} \to \texttt{B4} \to \texttt{B5}$  
  Condition: $\mathtt{p0} \land \neg \mathtt{p1} \land \neg \mathtt{p2}$
  \item Path 4 ($\mathtt{x1}$):
  $\texttt{B0} \to \texttt{B4} \to \texttt{B5}$  
  Condition: $\neg \mathtt{p0}$
\end{itemize}

Thus, the gating-path expression selecting $\mathtt{x0}$ is:
\[
C_{\mathtt{x0}}
=
(\mathtt{p0} \land \mathtt{p1}) \lor
(\mathtt{p0} \land \neg\mathtt{p1} \land \mathtt{p2})
\]
\end{example}

By analyzing gating paths, we augment each $\phi$-function with the corresponding Boolean guard so that the selected value becomes explicit.
In GSA, this produces $\gamma$-instructions, of the form $x = \gamma(P, x_{\mathrm{true}}, x_{\mathrm{false}})$, 
meaning:
\begin{itemize}
\item if $P$ evaluates to true, then $x$ receives $x_{\mathrm{true}}$;
\item otherwise, $x$ receives $x_{\mathrm{false}}$.
\end{itemize}

\begin{example}
\label{ex_gamma_function}
In Figure~\ref{fig_ssa_to_gsa}(b), the $\phi$-function
$\mathtt{x2}=\phi(\mathtt{x0}:\texttt{B3},\ \mathtt{x1}:\texttt{B4})$ is replaced by the $\gamma$-instruction
$\mathtt{x2} = \gamma\bigl(\mathtt{p0} \land (\mathtt{p1} \lor \mathtt{p2}),
\mathtt{x0},
\mathtt{x1}\bigr)$.
Here, the guard
$\mathtt{p0} \land (\mathtt{p1} \lor \mathtt{p2})$
summarizes all the conditions under which control reaches \texttt{B5} via \texttt{B3}.
The negation of this condition covers the cases reaching \texttt{B4}.
Thus, the $\gamma$-function precisely captures the value selection previously implicit in SSA.
\end{example}

While $\mu$- and $\gamma$-instructions are explicitly constructed by the algorithm of \citet{tu1995}, $\eta$-instructions were introduced earlier by \citet{ottenstein1990} as part of the original definition of GSA.
An $\eta$-instruction such as $v_0 = \eta( v_{\mathit{exit}}, p)$ appears in a block that exits a loop where $v_{\mathit{exit}}$ is defined and whose iteration is controlled by predicate $p$.
Introducing such instructions is straightforward, as they naturally arise from the notion of loop-closed form.\footnote{In LLVM, the Loop-Closed SSA (LCSSA) form enforces that all values defined inside a loop are used only within that loop. If a value is needed outside the loop, LCSSA inserts a unique $\phi$-function at the loop exit to capture it. In our case, we augment this $\phi$-function with the predicate controlling the loop, as illustrated in Figure~\ref{fig_exampleSlices}(d).}

\subsection{Slice Identification}
\label{sub_identification}

Once a program has been converted into GSA form, we identify the set of variables that constitute its idempotent backward slice (Definition~\ref{def_slice}) via a backward traversal over the program’s dependence graph (Definition~\ref{def_dep_graph}).

Under GSA, all dependences relevant to slicing are explicit in the syntax of instructions. Thus, unlike classic slicing, there is no need to separately compute data and control dependences. For instance, an $\eta$-instruction
$x = \eta(p, x_{\mathit{exit}})$ introduces a control dependence $p \rightarrow x$ and a data dependence $x_{\mathit{exit}} \rightarrow x$. Similarly, a $\gamma$-instruction
$x = \gamma(P, x_{\mathrm{true}}, x_{\mathrm{false}})$ introduces two data dependences
$x_{\mathrm{true}} \rightarrow x$ and $x_{\mathrm{false}} \rightarrow x$, plus one control dependence $p \rightarrow x$ for every predicate $p \in P$.

\begin{example}
\label{ex_back_deps}
Figure~\ref{fig_ssa_to_gsa}(c) depicts the backward dependence graph for variable \texttt{x3}. Three incoming edges to \texttt{x2}---namely $\mathtt{p0} \rightarrow \mathtt{x2}$, $\mathtt{p1} \rightarrow \mathtt{x2}$, and $\mathtt{p2} \rightarrow \mathtt{x2}$---correspond to control dependences; all others are data dependences.
However, for slice extraction, this distinction is irrelevant: the backward traversal simply follows all edges.
\end{example}

The traversal adopts two stop criteria; that is, it terminates when either of the following conditions holds:
\begin{description}
\item[Loop:] A slice for criterion $v$ cannot leave the loop in which $v$ is defined.
\item[Function:] Slices are intra-procedural; they cannot cross function boundaries.
\end{description}

These conditions define the \emph{Stop Set} of a slice for criterion $v$: the collection of variables whose incoming edges do not need to be explored.
Function parameters are natural stop points.
They have no incoming dependencies; thus, they satisfy the \textbf{Function} condition. To enforce the \textbf{Loop} condition, we assign each GSA variable a \emph{loop depth}, equal to the loop nesting depth of its definition. The backward traversal for a criterion $v$ stops when it reaches either:
(i) a $\mu$-instruction defining a variable of the same loop depth as $v$, or
(ii) any variable of smaller depth.
This ensures that the slice remains local to the loop that defines $v$.

\begin{figure}[htb]
  \centering
  \includegraphics[width=1.0\textwidth]{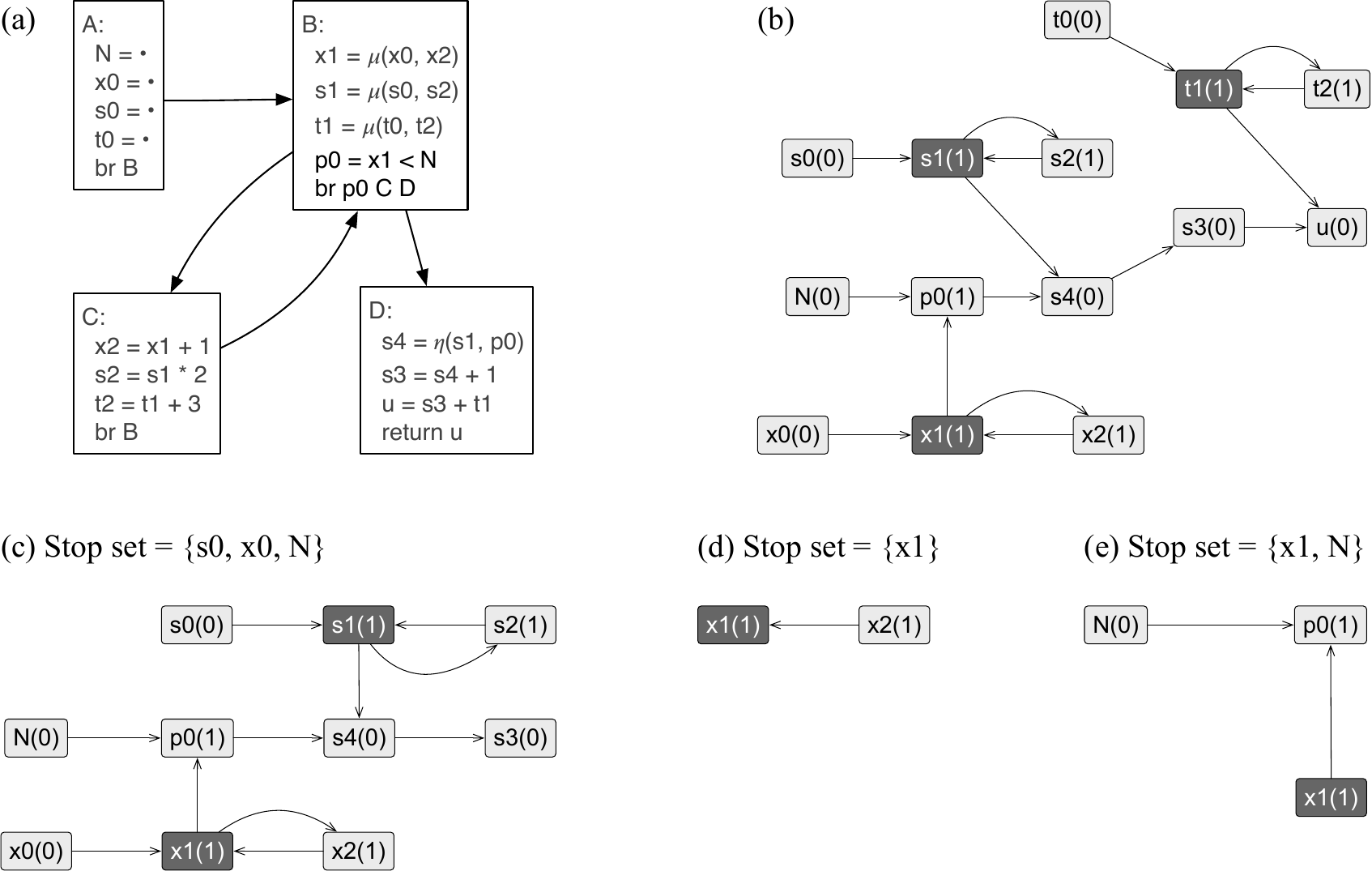}
  \caption{(a) GSA program (reproduced from Figure~\ref{fig_exampleSlices}). 
  (b) Dependence graph derived from def-use chains with loop depths.
  Variables defined by $\mu$-functions are marked in dark gray: they work as stop criteria.
  (c--e) Backward dependence graphs for \texttt{s3}, \texttt{x2}, and \texttt{p0}, respectively.}
  \Description{Examples of backward dependence graphs with stop sets.}
  \label{fig_dependencyGraphs}
\end{figure}

\begin{example}
\label{ex_stop_set}
Figure~\ref{fig_dependencyGraphs} shows the backward dependence graphs for variables \texttt{s3}, \texttt{x2}, and \texttt{p0} (previously seen in Figure~\ref{fig_exampleSlices}(d)).
Figure~\ref{fig_dependencyGraphs}(a) reproduces the GSA program; part (b) annotates each variable with its loop depth.  
Part (c) shows the dependence graph for \texttt{s3}, which has depth~0. The traversal reaches only function parameters, which therefore form the stop set.  
Part (d) shows the graph for \texttt{x2} (depth~1). The traversal stops at \texttt{x1}, defined by a $\mu$-instruction at the same depth.  
Finally, part (e) shows the graph for \texttt{p0} (also depth~1). The traversal stops both at the depth-1 definition of \texttt{x1} and at \texttt{N}, a depth-0 parameter.
\end{example}

Definition~\ref{def_dep_graph}, yields the complete and correct set of variables that integrate the backward slice, given that each node represents a program statement or SSA variable, and each edge denotes either a data or a control dependence.

\section{Slice-Based Code-Size Reduction}
\label{sec_size}

We call Slice-Based Code-Size Reduction (\texttt{SBCR}) a compiler transformation that applies the following sequence of steps on a program $P$:
\begin{enumerate}
\item Identify a set $S$ of idempotent backward slices in $P$ (see Section~\ref{sec_slices}); 

\item Outline these slices into a set $F$ of functions (see Section~\ref{sub_outlining});

\item Identify a subset $F' \subseteq F$ of functions that compute the same slice (see Section~\ref{sub_common_slice_identification});

\item Merge the functions in $F'$, and replace their slices with a single call
given a cost model that estimates code-size reduction
(see Section~\ref{sub_cost_model}).
\end{enumerate}

\subsection{Function Outlining}
\label{sub_outlining}

Once a slice has been identified, it can be extracted into a separate function through outlining.
This transformation is guarded by a set of legality checks that ensure the slice is idempotent; that is, that it causes no side effects and only reads from memory.
If all checks succeed, outlining proceeds in four steps:
\begin{enumerate}
    \item \textbf{Region Identification:} The set of basic blocks ($V_s$) identified by the backward traversal defines the slice region to be extracted (Definition~\ref{def_region}).
    \item \textbf{Interface Definition:} The function interface is created. Its parameters are the elements of the slice’s stop set, and its return type matches the type of the slicing criterion $v$.
    \item \textbf{Code Cloning:} The basic blocks in the region are cloned into a new function. A value map is used to remap references: internal dependencies are redirected to the cloned instructions, and external dependencies are replaced with the new function parameters.
    \item \textbf{Instruction Replacement:} The original region in the parent function is removed and replaced with a single \texttt{call} to the newly outlined slice function.
\end{enumerate}

\begin{definition}[Slice Region]
\label{def_region}
Let $G = (V, E)$ be a program’s control-flow graph, and let $S$ be the set of variables in the backward dependence graph for a slice criterion $v$.
The \emph{Slice Region} $V_s \subseteq V$ is the set of basic blocks that contain at least one instruction defining a variable in $S$.
\end{definition}

\paragraph{Recovering Control Flow in the Slice}
Reconstructing the control-flow graph of the slice requires rewriting branch instructions, because control-flow paths in the parent CFG may cross blocks outside $V_s$.  
The reconstruction relies on the notion of \emph{first dominator}, introduced by \citet{campos2023} and revisited below.

\begin{definition}[First Dominator]
\label{def_first_dom}
Given a control-flow graph $G = (V, E)$, a subset $V' \subseteq V$, and a node $b_0 \in V'$, the \emph{first dominator} of $b_0$ in $V'$ is the unique node $b_1 \in V'$ such that:
(i) $b_1$ dominates $b_0$ in $G$, and
(ii) for every other $b_2 \in V'$, $b_2 \neq b_1$, if $b_2$ dominates $b_0$, then $b_2$ also dominates $b_1$.
Operationally, $b_1$ is found via a backward walk along $G$’s dominator tree starting from $b_0$ and stopping at the first node in $V'$.
\end{definition}

While \citeauthor{campos2023} rely on this concept, they do not show that a slice region always contains a first dominator.
In our case, this property follows directly from Lemma~\ref{lem:entry-dominator}\footnote{Proofs of lemmas and theorems are available in Section 4.2 of the first author's master dissertation~\cite{Alvarenga25}}.

\begin{lemma}[Single-Entry Region]
\label{lem:entry-dominator}
Let an Idempotent Backward Slice $S$ be constructed by a backward dependence traversal from a criterion $v$, identifying a slice region $V_s$.
Then, there exists a unique block $b_{\mathit{entry}} \in V_s$ that dominates every other block in $V_s$.
\end{lemma}

Given a CFG $G = (V, E)$ and a slice region $V_s \subseteq V$, the slice’s CFG edges $E_s$ are reconstructed using two rules:
\begin{description}
\item[Transposition:] If $b_0, b_1 \in V_s$ and $b_0 \rightarrow b_1 \in E$, then $b_0 \rightarrow b_1 \in E_s$.
\item[Attraction:] If $b_1 \in V_s$, $b_0 \rightarrow b_1 \in E$, and $b_0 \notin V_s$, then let $b$ be the first dominator of $b_0$ in $V_s$; insert $b \rightarrow b_1$ into $E_s$.
\end{description}

\begin{figure}[htb]
  \centering
  \includegraphics[width=1.0\textwidth]{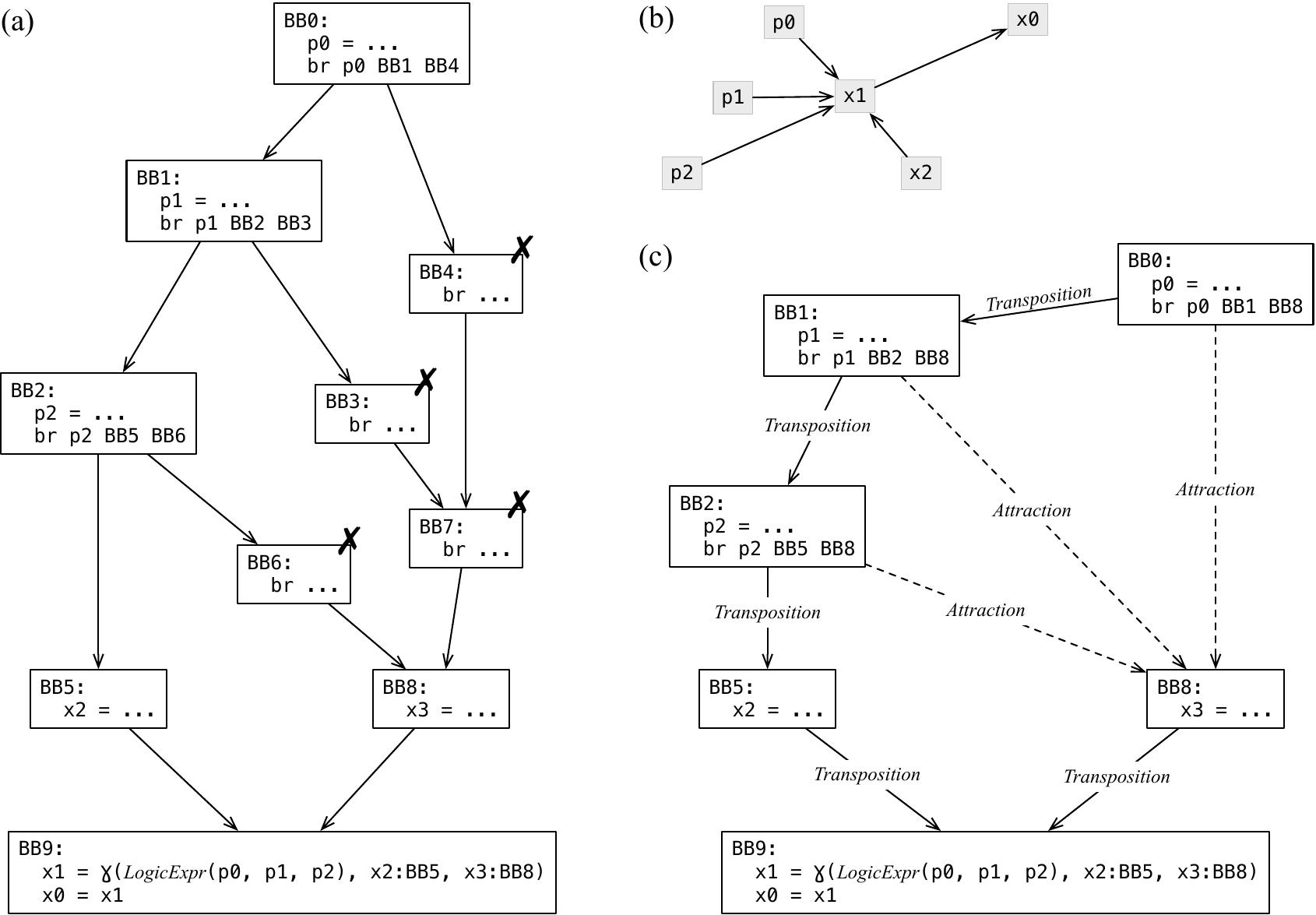}
  \caption{(a) A program that contains the common ``ladder'' CFG pattern.
  (b) Backward dependence graph for criterion \texttt{x1}.
  (c) Reconstructed CFG via transposition and attraction.}
  \Description{Examples of CFG reconstruction for slice extraction.}
  \label{fig_attractors}
\end{figure}

\begin{example}
\label{ex_attractors}
Figure~\ref{fig_attractors} illustrates CFG reconstruction using transposition and attraction.
The slice region is $V_s = \{\mathtt{BB0}, \mathtt{BB1}, \mathtt{BB2}, \mathtt{BB5}, \mathtt{BB9}\}$.
Edges between nodes inside $V_s$ are preserved directly (e.g., $\mathtt{BB5} \rightarrow \mathtt{BB9}$).
Other edges must be redirected.
For example, $\mathtt{BB6} \rightarrow \mathtt{BB8}$ is removed because $\mathtt{BB6} \notin V_s$; it is replaced with $\mathtt{BB2} \rightarrow \mathtt{BB8}$ since $\mathtt{BB2}$ is the first dominator of $\mathtt{BB6}$ within $V_s$.
\end{example}

Slices produced by backward dependence traversal followed by transposition and attraction form single-entry functions and preserve the observable behavior of the original program over the slice criterion.
This is formalized by Theorem~\ref{th:slicing-soundness}.

\begin{theorem}[Semantic Soundness]
\label{th:slicing-soundness}
Let $P$ be a program, $I_v$ be a slicing criterion defining variable $v$, and let $S = \mathrm{Slice}(P, I_v)$ be the extracted slice with single entry $L_{\mathit{entry}}$.
For any initial store $\sigma_{\mathit{in}}$:

\[
\text{If } 
P, I_v \vdash \langle L_{\mathit{entry}}, \sigma_{\mathit{in}} \rangle \rightarrow^* \sigma_P
\quad
\text{then } 
S, I_v \vdash \langle L_{\mathit{entry}}, \sigma_{\mathit{in}} \rangle \rightarrow^* \sigma_S
\quad \text{and} \quad
\sigma_S[v] = \sigma_P[v].
\]
\end{theorem}

\subsection{Common Slice Identification and Elimination}
\label{sub_common_slice_identification}

The identification and elimination of common slices leverages logic already implemented in the LLVM compiler. In our current prototype, we outline every slice whose criterion is a binary instruction. Thus, if a program contains $O(N)$ such instructions, the outliner may initially generate $O(N)$ slice functions, and run a worst case $O(N^2)$ operations, as stated in Theorem~\ref{thm:extraction_complexity}\footnote{Proof available in \citet[]{Alvarenga25}'s master dissertation.}.

\begin{theorem}
\label{thm:extraction_complexity}
Let $N$ be the number of instructions in a given program. If the number of candidate slice criteria is $O(N)$, and the worst-case time complexity to perform both backward dependence traversal and outlining for a single slice is $O(N)$, then the total worst-case time complexity for the complete slice extraction phase is $O(N^2)$.
\end{theorem}


Once all candidate slices have been outlined, we detect and eliminate duplicates using LLVM's standard \texttt{mergefunc} pass. This pass operates in two phases:
\begin{description}
\item[\textbf{Phase 1: Structural Hashing (Fast Filtering).}]  
Each outlined function is assigned a structural hash based on its instructions, constants, and control-flow skeleton. Functions with distinct hashes are immediately classified as different. This step reduces what would be a quadratic number of pairwise comparisons to near-linear grouping.
\item[\textbf{Phase 2: Canonical Comparison (Equivalence Checking).}]  
Within each hash group, a recursive equivalence test is performed. The pass attempts to build a bijection between arguments, instructions, and basic blocks to prove that two functions are identical modulo renaming. When equivalence is established, one implementation is selected as canonical and its duplicates are replaced with lightweight aliases or thunk calls.
\end{description}
After canonicalization, the cost model introduced in Section~\ref{sub_cost_model} determines which slice functions are profitable: only those predicted to reduce code size are retained.
Notice that if outlining a particular set of similar slices is not considered worth it, then the functions that represent them are simply removed from the program, and no modification is caused on the target code.
In other words, our implementation does not outline and then inline back: it rather removes the outlined functions deemed unprofitable.
Otherwise, the slicing criterion (the original defining instruction) is replaced with a call to the merged slice function, and LLVM’s \texttt{simplifyCFG} pass removes any remaining dead code in the parent function.

\subsection{Cost Model}
\label{sub_cost_model}

As with many compiler optimizations, the effectiveness of \texttt{SBCR} depends on a cost model.
In this paper, we use a cost model parameterized by three values $(I, P, C)$, defined as follows:
\begin{itemize}
\item $I$: the number of instructions that implement the outlined slice $S$;
\item $P$: the number of parameters of the function representing $S$;
\item $C$: the number of occurrences of $S$ in the target program. This number approximates the number of calls to the outlined function, if the slice is outlined and then merged with other outlined slices that are similar.
\end{itemize}
Based on these parameters, the optimization may either outline $S$ into a separate function or leave it in place.
If the slice is left in place, then it does not cause any modification in the code.
In our implementation, a slice $S$ is outlined only if $(3 < I \leq 20,\, P \leq 1,\, C \geq 10)$.
To determine these thresholds, we performed a search over the space $(I = \{10, 20, 40, 80, 160\},\, 0 \leq P \leq 20,\, C = \{10, 20, 40, 80, 160, 320, 640\})$, using the LLVM Test Suite as training data.
Section~\ref{sub_eval_cost_model} provides further details on this search.

\section{Evaluation}
\label{sec_eval}

This section empirically evaluates the slicing approach from Section~\ref{sec_size}, implemented as an \textit{out-of-tree} LLVM pass.
It analyzes the following research questions:

\begin{description}
\item[RQ1:] What are the ideal cost model parameters for \texttt{SBCR}?
\item[RQ2:] How much code-size reduction can idempotent slice outlining achieve, and how does it compare to previous work?
\item[RQ3:] What is slice outlining's impact on benchmark runtime, and how does it compare to previous work?
\item[RQ4:] What compilation overhead does slice outlining add, and how does it compare to previous work?
\item [RQ5:] What is the asymptotic behavior of the slice outlining algorithm?
\item [RQ6:] What is the time taken by the different phases of the proposed optimization?
\item [RQ7:] Is there a cumulative benefit to combining \citet{rocha2020}'s code merging, LLVM's outliner, and \texttt{SBCR}?
\end{description}

\paragraph{Benchmarks and Size Metrics}
Experiments run on all the \DaedalusCompTotal{} benchmarks available in the LLVM test suite (\url{https://github.com/llvm/llvm-test-suite}).
This suite includes 134 unit test programs, 1,548 regression test programs from the \texttt{gcc-c-torture}, C, and C++ collections, 307 programs from benchmark suites such as MiBench, PolyBench, TSVC, and others, and 18 programs from real-world applications.
The size of these programs is reported in two ways: using the number of LLVM instructions in the program's intermediate representation, and using the size of the \texttt{.text} section of the executable.
We confirm that, for this entire test suite, our implementation yields binaries that run correctly.

\paragraph{Baselines}
Henceforth, we shall refer to the technique proposed in this paper as \textbf{\texttt{SBCR}} (Slice-Based Code-Size Reduction).
We will compare it against two baselines from previous work:
\begin{description}
\item [\texttt{FMSA}:] The technique introduced by \citet{rocha2020} in 2020, which uses sequence alignment to identify common instructions, and merge them into a single control-flow path.

\item [\texttt{IROutliner}:] A standard pass available in LLVM, which identify sequences of recurrent instructions within basic blocks, and extracts them into common functions.
\end{description}

\paragraph{Hardware and Software}
Experiments were conducted on a dedicated server featuring an AMD CPU (Ryzen Threadripper 7970X) with 32-Cores at 4GHz and 128 GiB of RAM.
The operating system is Ubuntu 24.04.3 LTS (GNU/Linux 6.8.0-85-generic x86\_64).
Experiments run on a multi-threaded setting; however, this fact bears no effects on the results, besides speeding up the evaluation.

\paragraph{Compilation Pipeline}

\begin{figure}[ht]
  \centering
  \includegraphics[width=1.0\textwidth]{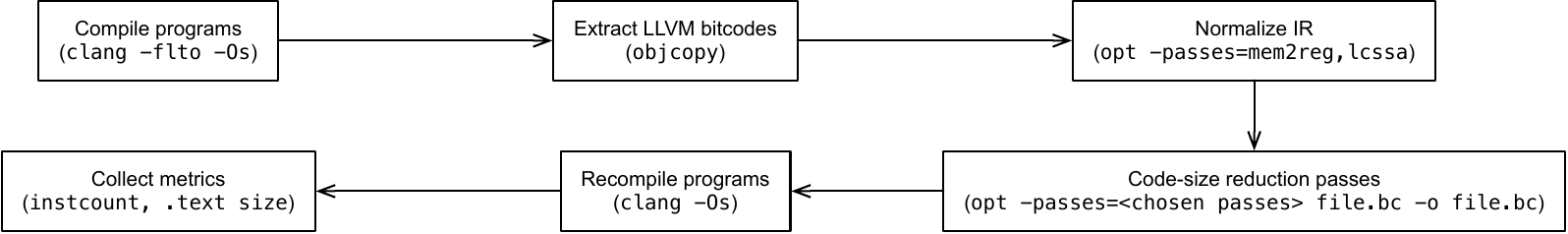}
  \caption{The compilation pipeline used to evaluate the different code-size reduction techniques. Notice that the code-size reduction approach is applied at link-time, onto the single bitcode file that contains all the modules of a multi-source benchmark. Every variation reported in this section is on top of \texttt{clang -Os}.}
  \Description{Examples of different program slices.}
  \label{fig_compilationPipeline}
\end{figure}

To isolate the impact of each optimization pass, we measured metrics by compiling every program with the \texttt{-Os} flag, both with and without the target pass applied. As this is not a default feature of the LLVM Test Suite, we modified its build system to create a post-build pipeline, following the pipeline in Figure~\ref{fig_compilationPipeline}.
This pipeline first compiles programs with Link Time Optimization (LTO) to embed bitcode into the executable. Then we use \texttt{objcopy} to extract the bitcode, normalize it with canonicalization passes (\texttt{mem2reg}, \texttt{lcssa}), apply the target optimization, and finally recompile the transformed bitcode with \texttt{-Os} to collect metrics.

\paragraph{Summary of Results}
The experimental results are summarized in Table~\ref{tab:overall-metrics}. For each optimization pass, the table reports the total number of programs affected by the transformation, and the overall geometric mean of the change. Positive percentages in the geometric mean indicate an increase in a metric, while negative percentages indicate a decrease. Notice that if we consider optimization along these four different axes: \texttt{Instcount}, \texttt{.text} segment size, execution time and compilation time, then none of the three transformations evaluated in this section is able to reduce them all.
However, if we focus on specific benchmarks, then we are able to see significant improvements in any of them, as we shall analyze in the rest of this section.

\begin{table}[htb]
    \centering
    \caption{Summary of results.
    \textbf{Affected}: the number of programs affected by the transformation, out of \DaedalusCompTotal{} programs.
    \textbf{Geom.:} the geometric mean of relative change. Negative percentages indicate a reduction in the metric, whereas positive percentages indicate an increase.}
    \label{tab:overall-metrics}
    \begin{tabularx}{\textwidth}{l*{9}{Y}}
        \toprule
        & \multicolumn{2}{c}{\textbf{SBCR}} & \multicolumn{2}{c}{\textbf{Func. Merging}} & \multicolumn{2}{c}{\textbf{IROutliner}} \\
        \cmidrule(lr){2-3}\cmidrule(lr){4-5}\cmidrule(lr){6-7}
        \textbf{Metric} & \textbf{Affected} & \textbf{Geom.} & \textbf{Affected} & \textbf{Geom.} & \textbf{Affected} & \textbf{Geom.} \\
        \midrule
        Instcount    & \subtractAndPrint{\DaedalusInstTotal}{\DaedalusInstUnchCount} & \DaedalusInstOverallGeo  & \subtractAndPrint{\FMInstTotal}{\FMInstUnchCount} & \FMInstOverallGeo  & \subtractAndPrint{\IROInstTotal}{\IROInstUnchCount} & \IROInstOverallGeo \\
        .text size    & \subtractAndPrint{\DaedalusSizeTotal}{\DaedalusSizeUnchCount} & \DaedalusSizeOverallGeo  & \subtractAndPrint{\FMSizeTotal}{\FMSizeUnchCount} & \FMSizeOverallGeo  & \subtractAndPrint{\IROSizeTotal}{\IROSizeUnchCount} & \IROSizeOverallGeo \\
        Exec. Time   & \subtractAndPrint{\DaedalusExecTotal}{\DaedalusExecUnchCount} & \DaedalusExecOverallGeo  & \subtractAndPrint{\FMExecTotal}{\FMExecUnchCount} & \FMExecOverallGeo  & \subtractAndPrint{\IROExecTotal}{\IROExecUnchCount} & \IROExecOverallGeo \\
        Comp. Time & \subtractAndPrint{\DaedalusCompTotal}{\DaedalusCompUnchCount} & \DaedalusCompOverallGeo  & \subtractAndPrint{\FMCompTotal}{\FMCompUnchCount} & \FMCompOverallGeo  & \subtractAndPrint{\IROCompTotal}{\IROCompUnchCount} & \IROCompOverallGeo \\
        \bottomrule
    \end{tabularx}
    \begin{center}
    \end{center}
\end{table}

The remainder of this section analyzes each of the proposed research questions. Table~\ref{tab:overall-metrics} summarizes the central findings. The main conclusions drawn from our evaluation are listed below:
\begin{description}
\item[RQ1:] The most effective \texttt{SBCR} configuration outlines slices that have at most one parameter, contain fewer than 20 LLVM instructions, and occur at least 10 times in the program.
\item[RQ2:] Neither \texttt{SBCR} nor \texttt{FMSA} reduces \texttt{.text} size universally across all benchmarks; however, when restricted to programs where they are beneficial, both approaches achieve average reductions close to $-10\%$.
\item[RQ3:] \texttt{SBCR} does not introduce statistically significant performance overhead in runtime execution.
\item[RQ4:] \texttt{SBCR} increases overall compilation time by approximately $4\%$ on average.
\item[RQ5:] In practice, the execution time of \texttt{SBCR} scales linearly with program size, when size is measured as the number of LLVM IR instructions.
\item[RQ6:] The most computationally expensive component of \texttt{SBCR} is the slice identification phase.
\item[RQ7:] Combining all three code-size reduction techniques yields larger reductions than any technique applied individually.
\end{description}

\subsection{RQ1: Cost Model}
\label{sub_eval_cost_model}

As described in Section~\ref{sub_cost_model}, the decision of whether a slice should be outlined is guided by a three-parameter cost model. To determine effective parameter values for code-size reduction, we performed an exhaustive search over a bounded design space, using the LLVM Test Suite as our evaluation corpus. The search consisted of 735 compilation runs, varying the following thresholds: maximum number of parameters \mbox{($0 \leq P \leq 20$)}, maximum number of instructions in the slice \mbox{($I \in \{10,20,40,80,160\}$)}, and minimum number of slice occurrences \mbox{($C \in \{10,20,40,80,160,320,640\}$)}.

\begin{figure}[ht]
    \centering
    \includegraphics[width=1\textwidth]{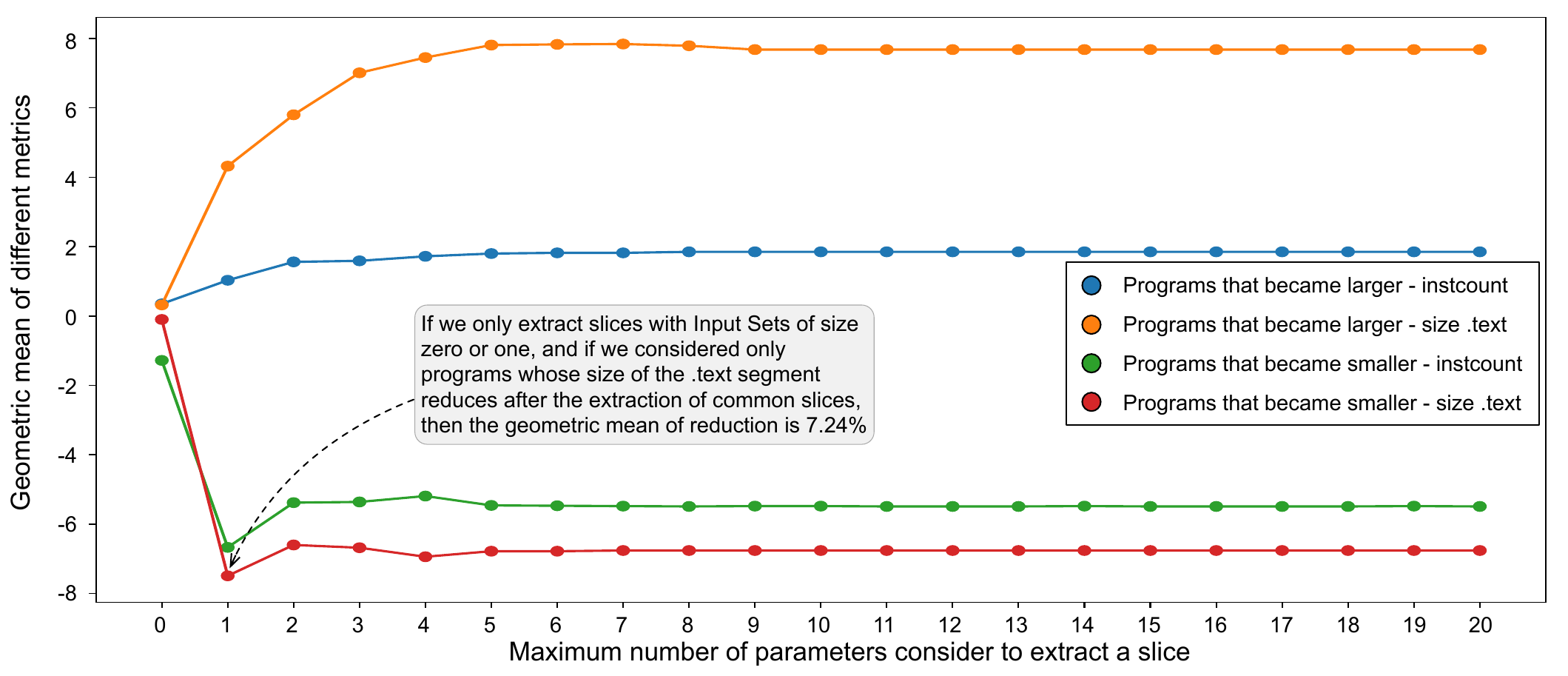}
    \caption{Impact of the maximum number of slice parameters on the geometric mean of code-size metrics.}
    \Description{Impact of the maximum number of parameters on code-size reduction.}
    \label{fig_graph_Params}
\end{figure}

\paragraph{Discussion.}
Figure~\ref{fig_graph_Params} shows the geometric mean of the reduction in code size while varying the allowed number of function parameters. The results indicate that the best reductions occur when the Input Set of a slice is very small—preferably containing at most one variable. These variables become the formal parameters of the outlined function.

While the optimal choice of parameter $P$ is dependent on the thresholds for $I$ (instruction count) and $C$ (call count), our evaluation consistently favored configurations where $P \leq 1$. The number of function parameters significantly affects whether the function's prolog and epilogue will introduce considerable overhead. The best measured reduction in \texttt{Instcount} was achieved when outlined slices were limited to at most $I = 20$ instructions and occurred at least $C = 10$ times within the program.

\subsection{RQ2: Code-Size Reduction}
\label{eval_eval_code_reduction}

This section investigates the effects of \texttt{SBCR} (and previous work) on the LLVM test suite.
The methodology compiles each program using the pipeline in Figure~\ref{fig_compilationPipeline} and records its \texttt{.text} segment size and instruction count (\texttt{Instcount} metric).

\paragraph{Discussion}
If used without restrictions, then the slice-based approach will result in a slight geometric mean increase of \text{\DaedalusSizeOverallGeo} across the \DaedalusSizeTotal~programs.
However, if restricted only to programs where it can lead to code reduction (on both \texttt{Instcount} and \texttt{.text} size metrics), then it causes an overall reduction of the \texttt{.text} segment of \text{\NegInstTextDaedalusB} programs with a geometric mean of reduction of \text{\DiffGeoNegInstDaedalusB}, and \text{\DiffGeoNegInstcountDaedalusB} in \texttt{Instcount} metric). If we analyze only the subset of programs that reduces only the \texttt{Instcount} metric, then for \text{\DaedalusSizeLowCount} programs, the geometric mean of reduction is \text{\DaedalusSizeLowGeo} programs.
Notice that although the three approaches were designed to reduce code, they might lead to code increase due to different reasons:
\begin{itemize}
\item The slice-based approach and function merging by sequence alignment might lead to code increase due to the duplication of control-flow instructions, namely conditional and unconditional branches.
Even though the extraction of slices lead to dead-code elimination on the parent function, it might still leave branches untouched, as these branches target basic blocks containing instructions that were not part of the original slice.
Thus, these branches will remain in the parent function and in the outlined slice.
\item The LLVM IR outliner targets only sequences of instructions without branches.
However, the extra instructions necessary to convert copy parameters to the outlined function, and then copy its return value back to the parent function might lead to code increase.
\end{itemize}

\begin{figure}[ht]
    \centering
    \includegraphics[width=1\textwidth]{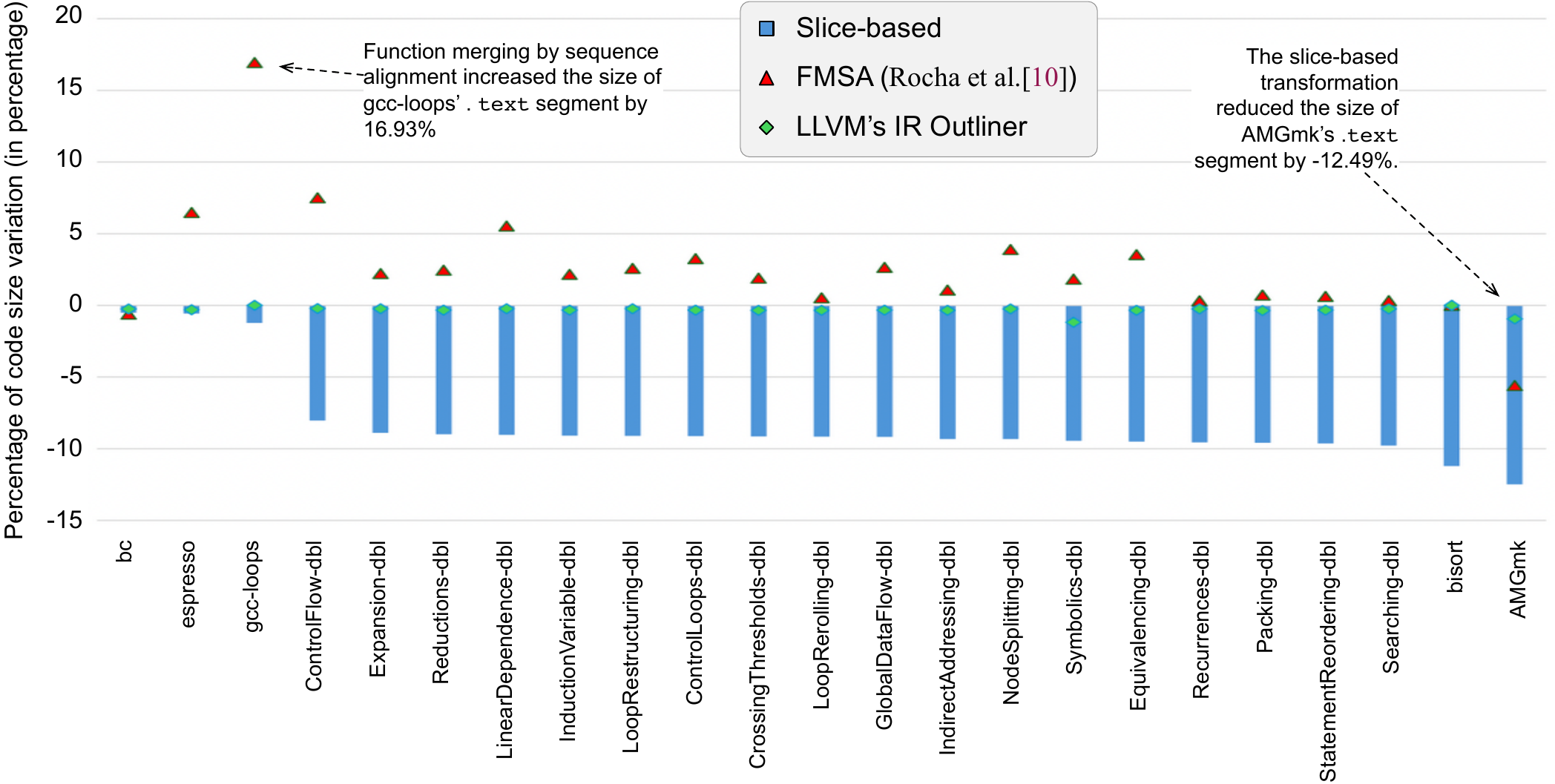} 
    \caption{Comparison between the different approaches on the 23 benchmarks where the slice-based technique got the largest percentage reductions.}
    \Description{Comparison between the different approaches on the 23 benchmarks where the slice-based technique got the largest percentage reductions.}
    \label{fig_daedalus_vs_baselines_reduced}
\end{figure}

\begin{figure}[ht]
    \centering
    \includegraphics[width=1\textwidth]{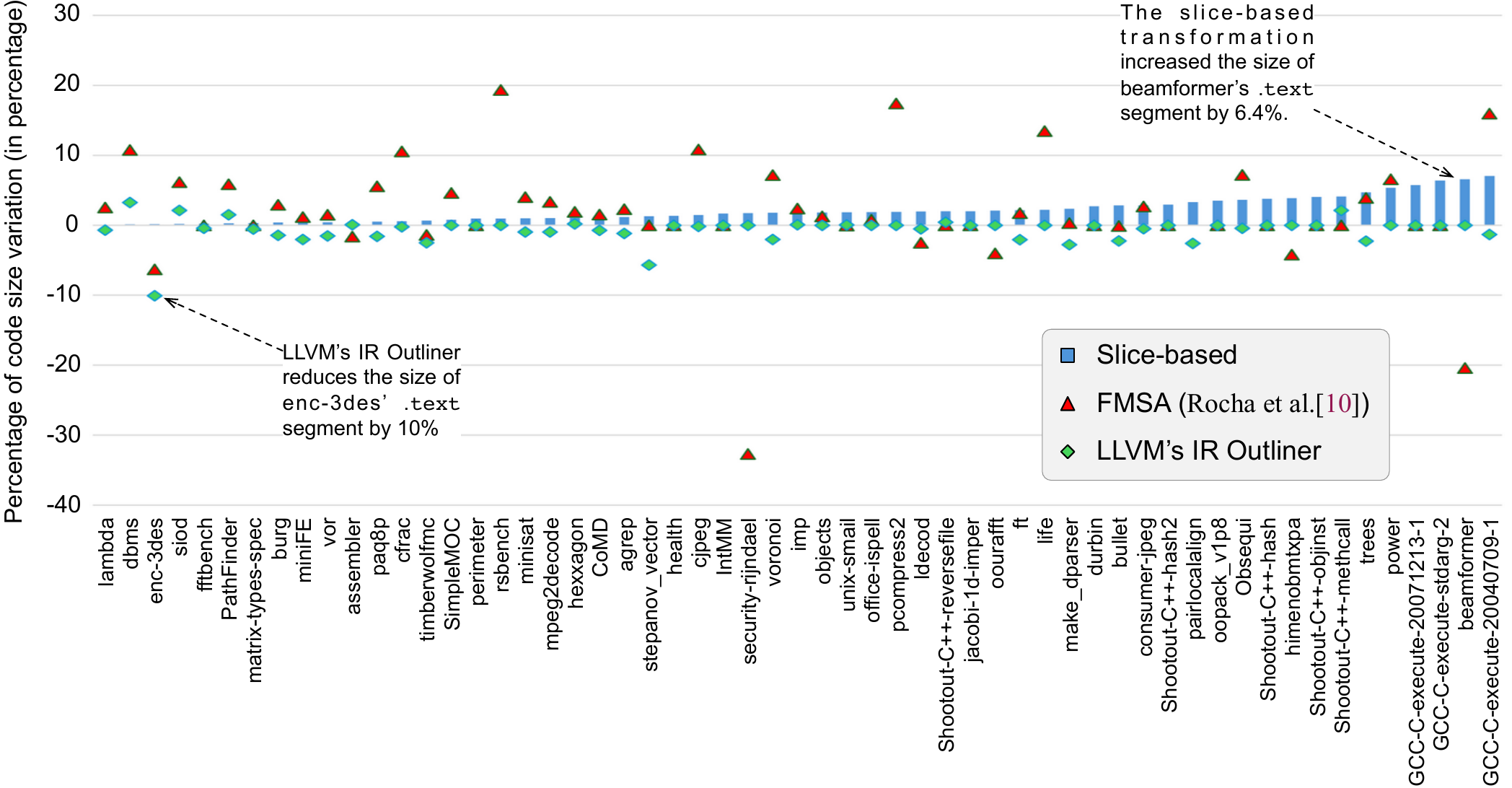} 
    \caption{Comparison between the different approaches on the 57 benchmarks where the slice-based technique got the largest percentage increases.}
    \Description{Comparison between the different approaches on the 57 benchmarks where the slice-based technique got the largest percentage increases.}
    \label{fig_daedalus_vs_baselines_increased}
\end{figure}

\paragraph{No Technique Subsumes the Others}
In the cases where the slice-based technique reduces code, it significantly outperforms baselines. Figure~\ref{fig_daedalus_vs_baselines_reduced} shows this performance for the \text{\NegInstTextDaedalusB~programs} where our approach reduced both \texttt{Instcount} and \texttt{.text size}. In these cases, our technique (blue) provides substantial reductions, while \texttt{IROutliner} (green) is minimal and \texttt{FMSA} (red) often increases size.
For example, in \texttt{IndirectAddressing-dbl.test}, \texttt{SBCR} achieved a \text{\DaedalusSizeTextValue} reduction. \texttt{IROut\-liner} got \text{\IROutlinerSizeTextValue}, and \texttt{FMSA} increased size by \text{\FuncMergSizeTextValue}.

Figure~\ref{fig_daedalus_vs_baselines_reduced} demonstrates that \texttt{SBCR} finds redundancies in optimized code that the other algorithms miss; however, the opposite also happens.
Figure~\ref{fig_daedalus_vs_baselines_increased} compares the impacts for the \text{\PosInstTextDaedalus~programs} where our approach increased both \texttt{.text} size and LLVM IR instruction count.
In \texttt{ldecod.test}, a clear example of this trade-off, the slice-based reduction increased code size by \text{\DaedalusSizeTextValueInc}.
The \texttt{IROutliner}, in turn, achieved a small \text{\IROutlinerSizeTextValueInc} reduction, while \texttt{FMSA} yielded a significant \text{\FuncMergSizeTextValueInc} reduction.

\paragraph{A Deeper Look into a Worst-Case Benchmark}
Table~\ref{tab:activate_sps_comparison} shows details of one benchmark, \texttt{activate\_} \texttt{sps}, where \texttt{SBCR} caused non-negligible code-size increase.
\texttt{SBCR} increased \texttt{activate\_sps} from \OriginalLdecod\ to \DaedalusLdecod\ bytes (1.50\%). Growth happens because \texttt{SBCR} creates two small slice functions, each called twice.
The overhead from these fine-grained, non-recurrent patterns dominated savings.
\texttt{IROutliner} also increases the size of this same function to \IROutlinerLdecod\ bytes (6.42\%). It found two larger, recurrent regions (\CreatedFuncAIROLdecod and \CreatedFuncBIROLdecod\ bytes; \IRONewFuncsNumCalls\ total calls), but call-site complexity prevents a net reduction.
In contrast, \texttt{FMSA} reduces the function to \FMLdecod\ bytes (-1.46\%). It replaces code with \FMNewFuncsNumCalls\ calls to six merged functions (\CreatedFuncAFMLdecod–\CreatedFuncFFMLdecod\ bytes). Here, coarse-grained, recurrent transformations amortized overhead and yielded the best reduction.

\begin{table}[htbp]
\centering
\caption{Comparison of Code-Size Impact for \texttt{activate\_sps} Across Passes.}
\label{tab:activate_sps_comparison}
\resizebox{\textwidth}{!}{%
\begin{tabular}{lcccccc}
\hline
\textbf{Pass} &
\textbf{Final Size (B)} &
\textbf{$\Delta$ (B)} &
\textbf{\# Fns} &
\textbf{New Fn Sizes (B)} &
\textbf{\# Calls} &
\textbf{Granularity / Recurrence} \\
\hline
Original &
\OriginalLdecod &
0 &
0 &
-- &
-- &
Baseline \\

SBCR &
\DaedalusLdecod &
$\DaedalusLdecod - \OriginalLdecod$ &
2 &
\CreatedFuncADaedLdecod, \CreatedFuncBDaedLdecod &
\DaedNewFuncsNumCalls &
Fine / Low \\

IROutliner &
\IROutlinerLdecod &
$\IROutlinerLdecod - \OriginalLdecod$ &
2 &
\CreatedFuncAIROLdecod, \CreatedFuncBIROLdecod &
\IRONewFuncsNumCalls &
Medium / Moderate \\

FMSA &
\FMLdecod &
$\FMLdecod - \OriginalLdecod$ &
6 &
\CreatedFuncAFMLdecod, \CreatedFuncBFMLdecod, \CreatedFuncCFMLdecod, \CreatedFuncDFMLdecod, \CreatedFuncEFMLdecod, \CreatedFuncFFMLdecod &
\FMNewFuncsNumCalls &
Coarse / High \\
\hline
\end{tabular}
}
\end{table}

\subsection{RQ3: Running Time}
\label{eval_eval_time}

The three optimizations discussed in this section were primarily designed for code-size reduction rather than performance improvement. Consequently, their impact on execution time is generally limited: across most transformed programs, we observed no statistically significant differences. However, individual benchmarks do exhibit notable speedups or regressions, depending on the applied transformation. This section analyzes these variations in detail.

\paragraph{Discussion.}
Across all benchmarks, \texttt{SBCR}'s effect on execution time is not statistically significant, with a geometric mean runtime variation of \text{\DaedalusExecOverallGeo} (Table~\ref{tab:overall-metrics}). Even if we restrict observations to the programs that were effectively transformed by \texttt{SBCR}, the vast majority of them showed no measurable performance difference.
Where performance degraded, \texttt{SBCR} increased runtime by \DiffGeoPosExecDaedalus~across \NegInstPosExecDaedalus~programs. This increase is due to function call overhead. Similarly, \texttt{FMSA} and \texttt{IROutliner} saw increases of \DiffGeoPosExecFuncMerging~and \DiffGeoPosExecIROutliner~in \NegInstPosExecFuncMerging~and \NegInstPosExecIROutliner~programs, respectively.

We have also observed situations where the performance of the transformed benchmarks improved. \texttt{SBCR} reduced execution time by \DiffGeoExecDaedalus~in \NegInstExecDaedalus~benchmarks, while \texttt{FMSA} achieved a \DiffGeoExecFuncMerging~reduction in \NegInstExecFuncMerging~benchmarks, and \texttt{IROutliner} a \DiffGeoExecIROutliner~reduction in \NegInstExecIROutliner~benchmarks.
We believe that such improvements are due to better instruction cache locality, as we discuss in the rest of this section.

\paragraph{Instruction Cache Performance Analysis}
We analyzed hardware performance counters for \texttt{Global\-DataFlow-dbl}, which had the greatest reduction in execution time, instruction count, and \texttt{.text} size. Comparing the \texttt{-Os} baseline to the \texttt{SBCR} version reveals lower cache-miss rates, which we believe is due to improved instruction locality and cache utilization.
In the baseline configuration, hardware counters recorded \num{6027} branch loads and \num{2165} branch-load misses, alongside \num{1736} L1 instruction cache loads and \num{1163} misses. After applying \texttt{SBCR}, these values shifted to \num{5661} branch loads, \num{1870} branch-load misses, and \num{1011} L1 instruction cache loads with \num{600} misses.
This benchmark, \texttt{Global\-DataFlow-dbl}, is an ideal case for \texttt{SBCR}, as the optimization enhances both instruction and data locality. The L1 data cache miss ratio fell from $66.99\%$ to $59.35\%$. Control flow also became more predictable (branch miss ratio decreased from $35.92\%$ to $33.03\%$), showing improved memory efficiency without significant control-flow penalties.

\subsection{RQ4: Compilation Time}
\label{eval_eval_comp_time}

Slice-based code-size reduction, as defined in Section~\ref{sec_size}, is a relatively expensive transformation: it requires a backward dependence traversal for every binary instruction in a program, followed by the outlining of slices into standalone functions and the merging of equivalent functions. Nevertheless, this section shows that the associated overhead remains modest when compared to the overall cost of a full compilation pipeline.

\begin{figure}[ht]
    \centering
    \includegraphics[width=1\textwidth]{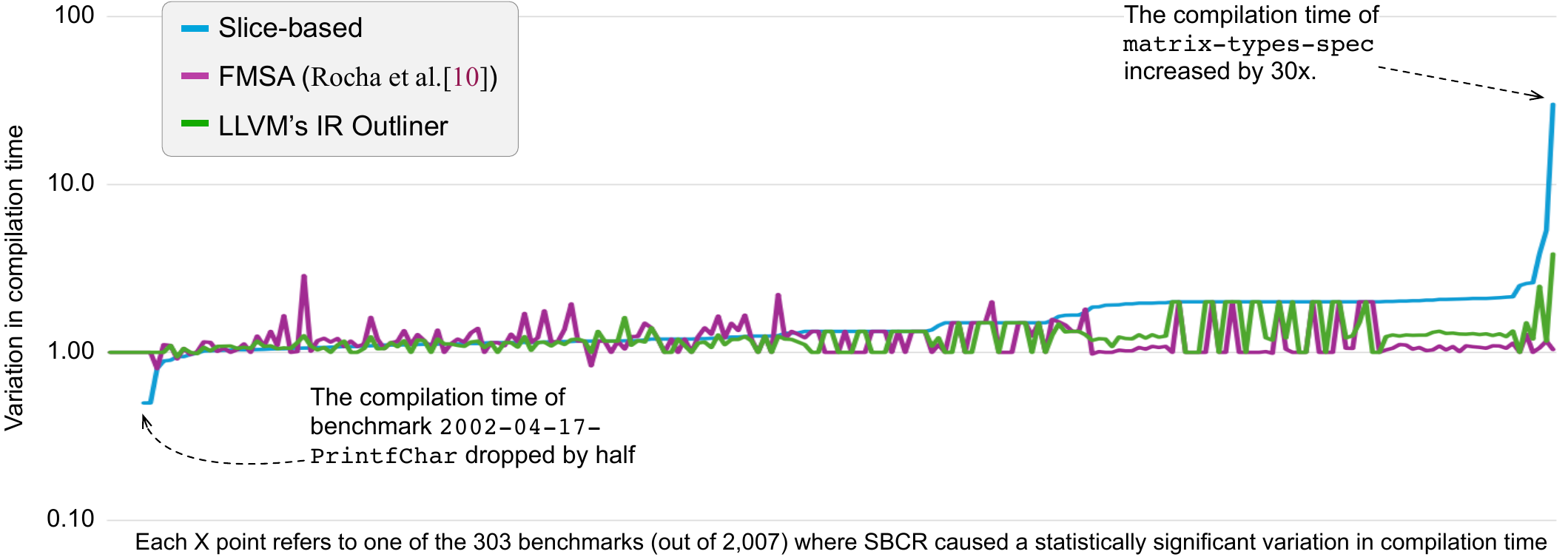}
    \caption{Variation in compilation time.}
    \Description{Variation in compilation time.}
    \label{fig_variation_compilation_time}
\end{figure}

\paragraph{Discussion.}
On average, \texttt{SBCR} increases compilation time by \text{\DaedalusCompOverallGeo}, computed as the geometric mean over the 303 programs whose compilation times changed (either positively or negatively).
Figure~\ref{fig_variation_compilation_time} shows how compilation time varies across this universe of benchmarks.
The remaining \text{\DaedalusCompUnchCount} programs exhibited no statistically significant variation in compilation time.
The two baselines we compare against also introduce compilation overhead, although to a lesser extent. \texttt{FMSA} increases compilation time by 2.06\% on the 241 programs that experienced changes (considering exclusively this optimization; i.e., this is not the universe of benchmarks analyzed in Figure~\ref{fig_variation_compilation_time}).
The \texttt{IROutliner}, in turn, increases compilation time by 2.48\% over 258 affected programs.

The dominant cost of \texttt{SBCR} is outlining, which involves cloning code into new functions and replacing references in the parent function. In contrast, constructing the GSA form and performing backward dependence discovery represent a small portion of the runtime. Section~\ref{eval_eval_phases} analyzes these phases.
Interestingly, in some benchmarks the compilation time improves despite the additional \texttt{SBCR} processing. In nine benchmarks with statistically significant code-size reductions, the geometric mean compilation time decreased by --19.81\%. These improvements stem from the reduced number of LLVM instructions produced by \texttt{SBCR}: since most LLVM analyses and transformations run in time proportional to program size, shrinking the IR can accelerate downstream passes.

\subsection{RQ5: Asymptotic Behavior}
\label{eval_eval_asymptotic}

Theorem~\ref{thm:extraction_complexity} states that \texttt{SBCR} has an $O(N^2)$ worst-case, where $N$ is the number of LLVM IR instructions in the program. However, this upper bound is rarely observed in practice. In this section, we provide evidence that \texttt{SBCR} exhibits near-linear behavior on real-world code. To this end, we selected the 100 largest programs in the test suite and measured the Pearson correlation between instruction count and the absolute execution time of each transformation.

\paragraph{Discussion.}
Figure~\ref{fig_plot_numtime_combined} shows regression lines for compilation time as a function of program size for the three code-size reduction techniques evaluated. The correlation between compilation time and instruction count is close to 1.0 in all cases: 0.84 for \texttt{SBCR}, 0.94 for \texttt{FMSA}, and 0.92 for \texttt{IROutliner}. This trend is visually corroborated by the scatter plots in Figures~\ref{fig_plot_numtime_combined}(a)--(c), where the points align closely to the fitted linear curves.
Although \texttt{SBCR} is theoretically quadratic, several practical properties contribute to its near-linear performance:

\begin{figure}[ht]
    \centering
    \includegraphics[width=1.0\textwidth]{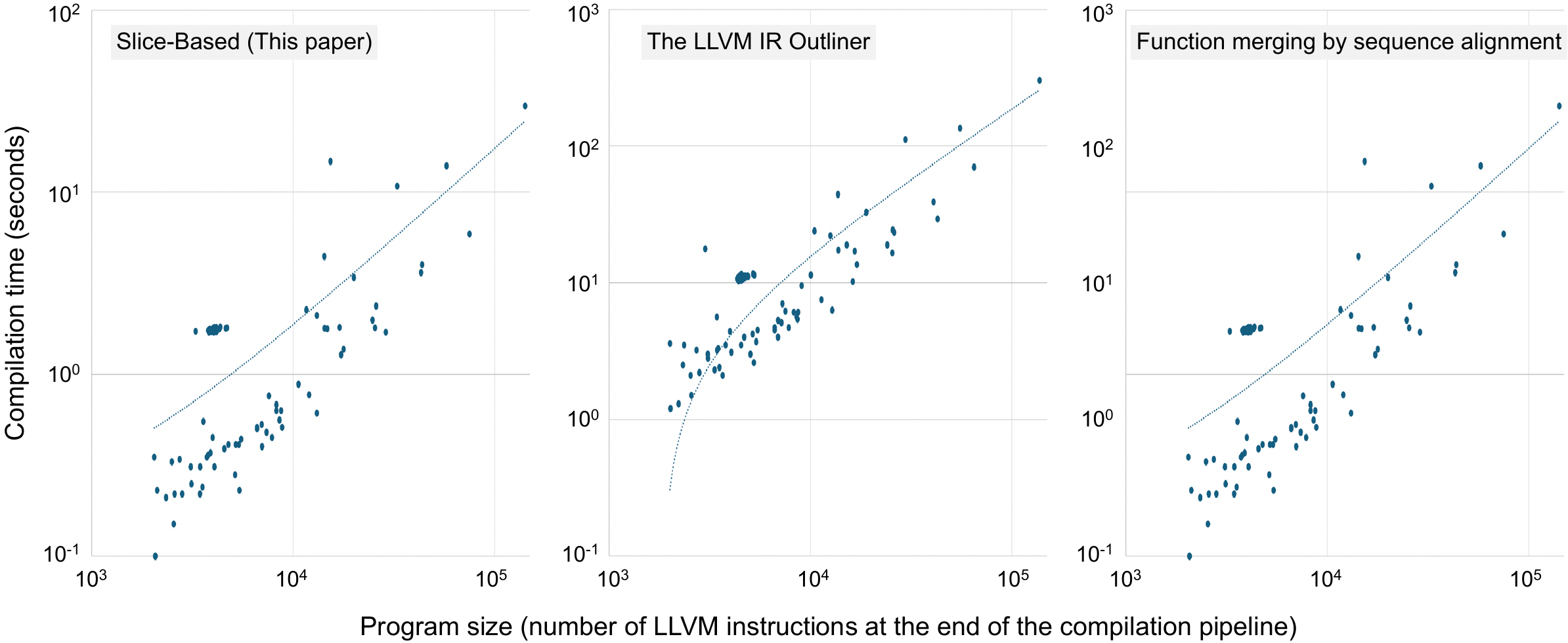}
    \caption{Correlation between compilation time and program size. Compilation time includes only the execution time of the code-size reduction pass. Program size is measured as the number of instructions at the end of the pipeline shown in Figure~\ref{fig_compilationPipeline}. Each point represents a benchmark. Axes are in logarithmic scale.}
    \Description{Correlation between compilation time and program size.}
    \label{fig_plot_numtime_combined}
\end{figure}

\begin{itemize}
\item \textbf{Most slices are small.} The backward dependency traversal (Section~\ref{sub_identification}) typically visits only a few instructions before reaching the slice criterion. Thus, this part of the algorithm behaves as $O(1)$ for most slices.
\item \textbf{Few slices satisfy the cost model.} Only a small number of candidates qualify for outlining and merging, making the interaction with LLVM’s \texttt{mergefunc} pass effectively constant—or at worst proportional to the number of accepted slices, which is typically low.
\end{itemize}
The remaining major component, GSA construction, is quasi-linear in the size of the program, which aligns with the linear correlations observed in Figure~\ref{fig_plot_numtime_combined}(a).

\subsection{RQ6: Time per Phase}
\label{eval_eval_phases}

In this section, we break down the transformation steps of \texttt{SBCR} and measure the time taken by each phase for a given input program.
We remind the reader that the \texttt{SBCR} pass is composed of four phases: Outlining, Merge, Remove Instructions, and Simplify, as Section~\ref{sec_size} explains. The \textit{Outlining Phase} contains procedures to construct the GSA form, identify program slices, check for valid slices to outline, and outline slices to new functions. The \textit{Merge} phase is responsible for merging all similar outlined functions. The \textit{Remove Instructions} phase is responsible for deleting instructions from the original function that become redundant after being moved into a newly merged function. Finally, all modified functions are simplified with the \texttt{simplifycfg} pass from LLVM.

\paragraph{Discussion}
To perform this study, we ran \texttt{SBCR} on all \text{\DaedalusCompTotal~tested} programs, collecting the percentage of execution time spent in each phase. We then computed the geometric mean of these percentages across all programs. The results are presented in Table~\ref{tab:phases-time}, which summarizes the main phases of \texttt{SBCR}, and Table~\ref{tab:subphases-timer}, which provides a detailed breakdown of the Outline phase.

\begin{table}[htbp]
\centering 
\begin{minipage}[t]{0.48\textwidth}
    \centering
    \caption{Timers for Outline Sub-Phases.}
    \label{tab:subphases-timer}
    \begin{tabular}{lc}
    \hline
    \textbf{Name} & \textbf{Wall Time} \\
    \hline
    Slice Identification Phase & 42.389\% \\
    GSA Construction Phase & 34.350\% \\
    canOutline Phase & 32.581\% \\
    Function Outline Phase & 11.541\% \\
    \hline
    \end{tabular}%
\end{minipage}%
\hfill
\begin{minipage}[t]{0.48\textwidth}
    \centering
    \caption{\texttt{SBCR} Phases (Outline includes Table ~\ref{tab:subphases-timer}).}
    \label{tab:phases-time}
    \begin{tabular}{lc}
    \hline
    \textbf{Name} & \textbf{Wall Time} \\
    \hline
    Outline Phase & 48.434\% \\
    Merge Phase & 22.885\% \\
    Remove Instructions Phase & 18.744\% \\
    Simplify Phase & 18.383\% \\
    \hline
    \end{tabular}%
\end{minipage}%
\end{table}

The column \textit{Name}, identifies the phase being measured, while column \textit{Wall Time} indicates the real elapsed time, including any waiting or synchronization delays. Percentages denote each phase's relative contribution to the overall compilation time.
Given the results in Tables~\ref{tab:subphases-timer} and~\ref{tab:phases-time}, we conclude that the \textit{Outline Phase} is the most time-consuming stage. Additionally, breaking down the \textit{Outline Phase}, \texttt{SBCR}'s \textit{Slice Identification} and \textit{GSA Construction} sub-phases expends significant time analyzing instruction's dependencies, and creating the mapping of $\phi$-functions' gates. The latter analysis is required to identify all data and control dependencies used by the former analysis.

\subsection{RQ7: Combined Results}
\label{eval_eval_combined}

We conducted an experiment to assess whether combining \texttt{SBCR}, \texttt{IROutliner}, and \texttt{FMSA} can achieve greater code-size reduction than running any of these optimizations individually. To this end, we evaluated all six permutations of the three passes across the entire LLVM Test Suite (\DaedalusCompTotal~programs) using the pipeline in Figure~\ref{fig_compilationPipeline}. We then selected the best-performing ordering and compared its results against applying each optimization alone.

\paragraph{Discussion.}
Table~\ref{tab:combination-metrics} summarizes the geometric means for each metric across the evaluated pass sequences. The best ordering was $\texttt{IROutliner} \rightarrow \texttt{SBCR} \rightarrow \texttt{FMSA}$.
This pipeline reduced the \texttt{.text} size in 86 benchmarks. Among these, the geometric mean reduction in instruction count was \(-14.43\%\), outperforming each of the three passes when used independently. It is worth noting that, if we restrict the comparison solely to \texttt{.text} size reductions, \texttt{FMSA} achieves slightly higher average reductions; however, it does so over a much smaller subset of programs (34 benchmarks).

\begin{table}[h!]
\centering
\caption{Effects of combined optimizations. Numbers in parentheses (first row) indicate the number of programs where \texttt{.text} size was reduced. Averages in each column refer to this subset.}
\label{tab:combination-metrics}
\begin{tabular}{lcccc}
\toprule
\textbf{Metric Name} &
\texttt{IROutliner-SBCR-FMSA} &
\texttt{IROutliner} &
\texttt{SBCR} &
\texttt{FMSA} \\
\midrule
\texttt{.text} size &
-9.68\% (86) &
-4.30\% (118) &
-7.24\% (29) &
-9.75\% (34) \\
\texttt{Instcount} &
-14.43\% &
-6.67\% &
-7.93\% &
-7.11\% \\
Exec. Time &
3.38\% &
-0.49\% &
4.55\% &
-0.05\% \\
Comp. Time &
39.85\% &
15.84\% &
65.10\% &
16.04\% \\
\bottomrule
\end{tabular}
\end{table}


\section{Related Work}
\label{sec_rw}

This work spans multiple areas in the programming languages literature, including classical program analyses and representations (e.g., GSA construction), as well as recent advances in code-size reduction. This section reviews key contributions in these domains.

\paragraph{Program Slicing.}
Program slicing is a longstanding research area originating from the seminal work of \citet{Weiser81}. As illustrated in Example~\ref{ex_exampleSlices}, our notion of \textit{idempotent slices} differs  from traditional program slices: rather than extracting subprograms, we construct referentially transparent functions. Consequently, slice identification and extraction operate on SSA values rather than program statements,as Section~\ref{sec_slices} explains.

In this sense, the work most closely related to ours is due to \citet{campos2023}, who use a variation of idempotent slices to transform call-by-value into call-by-name semantics. However, as discussed in Section~\ref{sub_problemWyvern}, both the definition of slices and the underlying algorithms are different. Their approach relies on \citet{Ferrante87}'s formulation of control dependencies, which they identify on the SSA graph using the sparse algorithm of \citet{Rodrigues16}. This approach may fail to recognize certain branches that influence slice results. Our technique avoids these limitations by leveraging the explicit dependency structure of GSA form.

\paragraph{GSA Construction.}
The Gated Static Single-Assignment form was introduced by \citet{ottenstein1990}, building on earlier work by \citet{Ferrante87} and emerging definitions of SSA form \cite{Alpern88,Cytron86,Rosen88}. The construction algorithm adopted in Section~\ref{sub_ssa_to_gsa} originates from \citet{tu1995}; however, our implementation follows the more recent formalization by \citet{herklotz2023}. \citeauthor{herklotz2023} provide a formal operational semantics for GSA-form programs and describe how to introduce gates into $\phi$-functions.
We opted to follow \citeauthor{herklotz2023}'s formulation because \citeauthor{tu1995} did not explicitly model $\eta$- and $\mu$-functions in the way required by our slice-based approach.
Nevertheless, none of this previous literature explores GSA form as a foundation for slicing. To the best of our knowledge, this work is the first to employ GSA form as a systematic basis for slice extraction.

\paragraph{Code-Size Reduction.}
Reducing code size by removing redundancy has been extensively studied, motivating a large body of recent research. Since instruction counts do not fluctuate statistically, code-size reduction has become a cornerstone of stochastic compiler optimization \cite{Faustino21a, Faustino21b, Armengol24}. Our approach, however, is entirely deterministic.
However, most previous deterministic instruction-outlining techniques focus on discovering common \textit{contiguous} instruction sequences \cite{Chabbi21}. In contrast, slice-based code-size reduction can identify non-contiguous regions containing branches or loops (as shown in Figure~\ref{fig_exampleSlices}).

Techniques such as \texttt{HyFM}~\cite{rocha_hyfm:_2021} and \texttt{FMSA}~\cite{rocha2020} are also capable of outlining non-sequential code. However, neither can merge redundant code located within the same function. The LLVM \texttt{IROutliner} \cite{Riddle2017Interprocedural} can do it, but it is not region-based; hence, cannot outline code spanning multiple blocks.
As demonstrated in Section~\ref{sec_eval}, \texttt{SBCR} is neither a subset nor a superset of \texttt{FMSA} or \texttt{IROutliner}. Like \texttt{IROutliner}, it can outline both intra- and inter-procedurally, but unlike either previous approach, it is \textit{region-based}. By identifying a slice criterion and including all instructions that it depends on, \texttt{SBCR} can extract more expressive and semantically coherent code regions, even across control-flow boundaries.

\section{Conclusion}
\label{sec_conclusion}

This paper introduces the concept of \emph{Idempotent Backward Slices} and presents a practical algorithm for identifying them in Gated SSA form. We argue that idempotent slices represent a new and expressive unit of redundancy: referentially transparent computations that can be extracted, duplicated, and invoked without changing program behavior. Based on this insight, we propose \texttt{SBCR}, a semantics-preserving transformation that outlines idempotent slices into single-entry functions, enabling precise and safe code-size reduction.
Our evaluation on more than 2{,}000 programs from the LLVM Test Suite demonstrates that:
\begin{itemize}
    \item \texttt{SBCR} achieves code-size reductions competitive with state-of-the-art approaches, especially in the presence of fine-grained redundancy.
    \item No single technique is universally superior: \texttt{SBCR}, \texttt{IROutliner}, and \texttt{FMSA} uncover distinct optimization opportunities.
    \item Although quadratic in theory, \texttt{SBCR} exhibits near-linear scaling in practice, making it viable in real-world compilation pipelines.
\end{itemize}

We view this work as a first step toward exploring idempotent slices as a general abstraction for redundancy elimination. A key challenge moving forward is preventing the transformation from increasing code size when applied aggressively. This concern motivates continued refinement of profitability heuristics and cost models, or integration with profile-guided optimization. These are directions that we leave open as future work.

\section*{Data Availability Statement}

The implementation of all the ideas discussed in this paper is publicly available at \url{https://github.com/lac-dcc/Daedalus}, under the GPL 3.0 license.
The experiments reported in Section~\ref{sec_eval} used LLVM~17, chosen for compatibility with the \texttt{FMSA}'s patch, \texttt{IROutliner}, and our pass.
We extended the \texttt{mergefunc} pass from LLVM to make possible the merging of functions within our pass.
The specific LLVM~17 build source code is available at \url{https://github.com/Casperento/llvm-project/tree/merge-functions-pass}.
The experiments were executed via shell scripts developed for this study, publicly available at \url{https://github.com/Casperento/daedalus-dbg-toolkit}.
We provide a docker image to reproduce all the experiments.
The \textit{Dockerfile} is located at \url{https://github.com/lac-dcc/Daedalus/tree/main/artifact/docker}.

\section*{Acknowledgment}

This project was supported by FAPEMIG (Grant APQ-00440-23), CNPq (Grant \#444127/2024-0), CAPES (\textsc{PrInt}), and Google, which sponsored part of Rafael Alvarenga's scholarship.  
We are grateful to Xinliang (David) Li and Victor Lee for their efforts in making the Google sponsorship possible.

\bibliographystyle{plainnat}
\bibliography{refs}

\end{document}